\documentclass[superscriptaddress,showpacs,amssymb,10pt,reprint,aps,prd,longbibliography,nofootinbib,floatfix]{revtex4-2}
\usepackage{graphicx,epsfig,amssymb} 
\usepackage{amsmath,amsfonts, times}
\usepackage{bm} 
\usepackage[normalem]{ulem}
\usepackage{epstopdf}
\usepackage{booktabs}
\usepackage{multirow}
\usepackage{array}
\usepackage{siunitx}
\usepackage{multirow}
\usepackage[linktocpage,colorlinks]{hyperref}
\usepackage[caption=false]{subfig}
\usepackage[usenames]{color}
\usepackage{mathrsfs}
\usepackage{natbib}
\usepackage{soul}
\usepackage{subfig}
\usepackage[utf8x]{inputenc}
\usepackage{tikz}
\usetikzlibrary{decorations.pathmorphing}

\definecolor{coolblack}{rgb}{0.0, 0.18, 0.39}
\definecolor{darkred}{rgb}{0.5,0,0}
\definecolor{darkgreen}{rgb}{0,0.5,0}
\definecolor{darkblue}{rgb}{0,0,0.5}
\definecolor{lapislazuli}{rgb}{0.15, 0.38, 0.61}
\definecolor{venetianred}{rgb}{0.78, 0.03, 0.08}
\definecolor{bleudefrance}{rgb}{0.19, 0.55, 0.91}
\definecolor{dogwoodrose}{rgb}{0.84, 0.09, 0.41}
\hypersetup{colorlinks=true, citecolor=darkgreen, linkcolor=darkblue,
urlcolor = blue}

\newcommand\numberthis{\addtocounter{equation}{1}\tag{\theequation}}

\begin{document}

\title{\large Non-minimal matter sector couplings in Lorentz-violating gravity: Self-consistent traversable wormholes and quasinormal modes}

\author{Renan B. Magalh\~aes}
    \email{renan.batalha@ufma.br}
	\affiliation{Programa de Pós-graduação em Física, Universidade Federal do Maranhão, Campus Universitário do Bacanga, São Luís (MA), 65080-805, Brazil.}
\author{Leandro A. Lessa}
	\email{leandrophys@gmail.com}
	\affiliation{Programa de Pós-graduação em Física, Universidade Federal do Maranhão, Campus Universitário do Bacanga, São Luís (MA), 65080-805, Brazil.}
\author{Rodolfo Casana}
	\email{rodolfo.casana@ufma.br}
	\affiliation{Programa de Pós-graduação em Física, Universidade Federal do Maranhão, Campus Universitário do Bacanga, São Luís (MA), 65080-805, Brazil.}
	\affiliation{Coordenação do Curso de  Física--Bacharelado, Universidade Federal do Maranhão, Campus Universitário do Bacanga, São Luís, Maranhão 65080-805, Brazil}

\begin{abstract}
The anisotropies induced by Lorentz-violating fields pose significant challenges for the search for compact objects in non-vacuum environments. In this work, nevertheless, we demonstrate that introducing couplings between Lorentz-violating fields and matter allows a remarkable class of spacetimes: traversable wormholes. Specifically, we consider additional couplings in the Lagrangian of a phantom scalar field and derive Ellis-Bronnikov spacetime analogs in a Lorentz-violating scenario where both a vector field and an antisymmetric rank-2 tensor field spontaneously acquire non-zero vacuum expectation values. Despite the distinct nature of these fields, their non-zero vacuum expectation values contribute additively to the overall effect on the phantom distribution and on the resulting line element.
Moreover, to probe the effects of the Lorentz violation in these spacetimes, we consider scalar perturbations lying in these spacetimes either coupled to the Lorentz-violating fields or minimally coupled to the metric. Notably, the additional Lorentz-violating couplings can alter scalar field dynamics so that perturbations propagate as if in a General Relativity background, thereby allowing for some traits of Lorentz violation to remain hidden. We compute the quasinormal mode spectra of these perturbations using three methods: direct integration, the 6th-order WKB approximation, and the Prony method, finding strong agreement among the results.
\end{abstract}
\date{\today}
\maketitle

\section{Introduction}
Einstein's theory has been confronted with theoretical and observational challenges that remain unresolved within its current framework. For instance, the observed flat rotation curves of galaxies~\cite{Bertone:2004pz}, the accelerated expansion of the universe~\cite{riess1998observational}, and the problem of singularities~\cite{hawking1970singularities}. Among these open questions, a complete quantum description of gravity is of particular importance because it is anticipated to become relevant at the Planck scale~\cite{rovelli2004quantum,weinstein2005quantum}. To address the issue of gravity quantization, several candidates for a quantum gravity theory have been proposed, including string theory~\cite{green2012superstring, gross1989heterotic}, loop quantum gravity~\cite{ashtekar1987new, rovelli1990loop}, causal dynamical triangulations~\cite{ambjorn2004emergence}, and Horava-Lifshitz gravity~\cite{hovrava2009quantum}. However, experimental verification of these theories remains a significant challenge, as direct experiments at the Planck scale are currently impractical. Nonetheless, suppressed effects originating from an underlying unified quantum gravity theory might be detectable through sensitive experiments at currently accessible low-energy scales using Effective Field Theories (EFT)~\cite{weinberg2016effective}. Among the potential observable signatures of an underlying unified quantum gravity theory are models incorporating Lorentz symmetry breaking~\cite{kostelecky1989spontaneous,liberati2013tests, kostelecky1995cpt,kostelecky1989phenomenological}.

Lorentz symmetry breaking can occur either explicitly or spontaneously~\cite{Kostelecky:2003fs}. Explicit Lorentz symmetry breaking (ELSB) is constructed through a Lagrangian density that includes terms no longer invariant under Lorentz transformations. Alternatively, it is possible to construct a theory where Lorentz symmetry is violated in a manner that preserves the independence of physical laws from specific reference frames, ensuring that the Lagrangian remains Lorentz invariant. The last occurs in the context of spontaneous Lorentz symmetry breaking (SLSB). In this framework, the trade-off is that the ground state of the physical system loses Lorentz symmetry. The core idea is that interactions among tensor fields in the underlying theory induce non-zero vacuum expectation values (VEVs) of some Lorentz tensors. An EFT for studying SLSB is the Standard-Model Extension (SME)~\cite{Kostelecky:2003fs,colladay1997cpt}. Furthermore, SLSB is the most suitable approach for application in the gravitational sector, as it ensures the conservation of the energy-momentum tensor~\cite{Bluhm:2007bd}. One of the simplest models exhibiting SLSB is the bumblebee model~\cite{Kostelecky:2003fs}, a dynamical vectorial theory with a self-interaction potential that triggers SLSB, extensively studied in the literature~\cite{bailey2006signals,capelo2015cosmological,maluf2013matter,maluf2014einstein,Lessa:2023yvw,maluf2021black}. Another candidate for SLSB is an antisymmetric rank-2 tensor field~\cite{Altschul:2009ae}, also known as the Kalb-Ramond field~\cite{kalb1974classical}, also endowed with a self-interaction potential that triggers SLSB. In both cases, if the tensor fields have non-vanishing VEVs, they define a preferred direction in the spacetime, signaling the violation of local Lorentz symmetry.

The anisotropies induced by the Lorentz-violating fields can remarkably modify the structure of compact objects, even in the absence of matter fields, as observed in Refs.~\cite{Casana:2017jkc, Ding:2019mal,  Maluf:2020kgf,filho2023vacuum,araujo2024exact,araujo2025does} for the bumblebee model and in Refs.~\cite{Yang:2023wtu, Liu:2024oas,araujo2025particle} for the antisymmetric rank-2 tensor model. Nevertheless, in non-vacuum environments, these anisotropies pose significant challenges for the search for compact objects~\cite{lessa2025self}. These difficulties mainly arise if the Lorentz-violating fields couple to gravity through non-minimal couplings solely, where the spacetime geometry becomes restricted by the field equations of the Lorentz-violating tensors in the VEV state. A suitable way to overcome these issues is to consider couplings between Lorentz-violating fields and matter, thereby modifying the field equations of the Lorentz-violating tensors and, in this manner, allowing the emergence of novel compact objects. By using such an approach in electro-vacuum scenarios, spherically symmetric charged black holes could be found~\cite{liu2025charged,duan2024electrically}. We remark that couplings between Lorentz-violating tensors and matter in the context of metric-affine bumblebee gravity have previously been investigated~\cite{delhom2022radiative,nascimento2024metric}.


Among the compact objects predicted by General Relativity (GR) and modified gravity theories, wormholes are undoubtedly considered one of the most intriguing~\cite{Visser:1995cc,de2025can,de2020general,lu2024gravitational,lu2025existence,di2017spin,battista2024generalized,de2021testing,de2021reconstructing,magalhaesCompactObjectsQuadratic2022,magalhaesAsymmetricWormholesPalatini2023,magalhaesEchoesBoundedUniverses2024}. These ultracompact objects (UCOs), with non-trivial topology, are characterized by a minimal hypersurface--known as throat--that serves as a bridge between two distant regions of the same or different universes. These tunnel-like structures may introduce novel features that tell wormhole apart from other UCOs. Initially, in the literature, wormholes were introduced as non-traversable bridges~\cite{flamm1916beitrage, einstein1935particle}. Nevertheless, in the 1970s, the Refs. \cite{ellisEtherFlowDrainhole1973, bronnikovScalartensorTheoryScalar1973, ellis1979evolving} found that phantom scalar fields could support traversable wormholes. The fact that the phantom field carries the ``wrong'' sign in front of its kinetic term leads to the violation of some energy conditions. Although this type of matter is considered exotic, phantom fields are gaining increasing attention in cosmology in recent years, as they give rise to an accelerated expansion of the Universe, i.e., a possible explanation for dark energy~\cite{caldwell2003phantom}. Moreover, recent results from the Dark Energy Spectroscopic Instrument (DESI) combined with Type Ia supernovae (SNIa) measurements~\cite{DES:2025bxy} may suggest a crossing of the phantom divide line. That implies that the dark energy equation-of-state parameter $w$ crosses the cosmological constant boundary $w_{\Lambda}=-1$. Theoretically, this suggests that within the framework of GR, phantom scalar fields would be required to consistently produce a phantom equation of state $w < -1$.

Until recently, only one traversable wormhole had been reported as a solution in the bumblebee model~\citep{ovgun2019exact}. However, that wormhole is not a self-consistent solution of the bumblebee gravity model, as demonstrated in Ref.~\cite{lessa2025self}. Thus, self-consistent traversable wormhole solutions in the bumblebee model have remained absent from the literature. Nevertheless, as shown in Ref.~\cite{Maluf:2021ywn}, it is possible to generate wormhole solutions with the VEV of a timelike antisymmetric rank-2 field by assuming an anisotropic matter source. Moreover, the counterpart of the Ellis-Bronnikov wormhole (EBWH) solution was found in Ref.~\cite{magalhaes2025wormholes} for the same VEV antisymmetric field non-minimally coupled to the Riemann tensor, assuming a phantom scalar field with a non-trivial self-interaction potential. Importantly, both wormhole solutions are self-consistent within the Kalb-Ramond framework.

In this work, we aim to derive an exact traversable wormhole solution in models featuring non-minimal couplings between the Lorentz-violating tensor fields (both the bumblebee and Kalb-Ramond fields) and the Ricci tensor, thereby filling a gap in the search for non-vacuum solutions in Lorentz-violating gravity models. We show that in order to generate such solutions, one must introduce couplings between the phantom scalar field and the Lorentz-violating fields, which we henceforth call LV-matter couplings, by following an approach similar to that introduced in Refs.~\cite{liu2025charged,duan2024electrically}. Thus, we derive EBWH analogs in a combined Lorentz-violating scenario incorporating both the bumblebee and the Kalb-Ramond fields. Despite the distinct nature of both fields, their non-zero VEVs contribute additively to the overall effect on the resulting line element and on the phantom distribution that is the source of this spacetime.

We remark that the Lorentz-violating fields can also influence an independent (canonical) scalar field lying in these Lorentz-violating spacetimes. The dynamics of a test scalar field with those additional couplings could be told apart from that of a test scalar field minimally coupled to the spacetime metric. Thus, the investigation of scalar waves either minimally coupled to the metric or coupled to the metric featuring additional LV-matter couplings is well-motivated, and one avenue for such exploration is the analysis of their distinct quasinormal mode (QNM) spectra. QNMs are represented by complex frequencies, with the real part corresponding to the oscillation frequency and the imaginary part determining the damping rate. Here, we calculate the QNM spectra of canonical scalar fields coupled to Lorentz-violating fields and minimally coupled to the metric using three independent methods: direct integration, the 6th-order WKB approximation, and the Prony method, finding strong agreement among the results. Above all, we find that the QNMs of scalar perturbations coupled to Lorentz-violating fields can coincide with the spectrum of scalar perturbations in the standard EBWH solution.

Our analysis reveals that such matter perturbations coupled to Lorentz-violating fields can propagate subject to an effective metric, which can even conceal the signatures of Lorentz violation, another novel result in the literature. We establish the necessary conditions for such a peculiarity to occur and show that this result, to find black hole solutions of Einstein-bumblebee gravity, can also be applied~\cite{Casana:2017jkc}.

The content of this paper is organized as follows. In Sec.~\ref{sec2}, we introduce the Lorentz-violating framework we will use, and derive its field equations. The wormhole solutions supported by a phantom scalar field coupled to Lorentz-violating fields are presented in Sec.~\ref{seciv}. In Sec.~\ref{sec:dynamics}, we investigate the dynamics of canonical scalar fields subjected either minimally coupled to the metric tensor or coupled to Lorentz-violating fields, calculating their distinct QNM spectra in Sec.~\ref{sec:qnm}. In Sec.~\ref{sec:a_note}, we examine the role of LV-matter coupling in the propagation of scalar perturbations within models featuring Lorentz symmetry breaking. Finally, we summarize our results and discuss some perspectives in Sec.~\ref{con}.

\section{Framework}\label{sec2}
Let us consider the gravity theory described by the action
\begin{equation} \label{ac1}
    S = \dfrac{1}{2\kappa}\int d^4x \sqrt{-g}\left(\mathcal{L}_{g}+\mathcal{L}_{K}+\mathcal{L}_{V}+\mathcal{L}_{M}\right),
\end{equation}
where $\mathcal{L}_{g}=\mathcal{L}_{EH}+\mathcal{L}_{LVg}$ is the gravitational Lagrangian, with $\mathcal{L}_{EH}=R$ being the Einstein-Hilbert term, where $R$ is the Ricci tensor and $g$ is the determinant of the metric tensor, and $\mathcal{L}_{LVg}$ encoding the  couplings of the bumblebee field, $\mathfrak{B}_{\mu}$, and the antisymmetric rank-2 tensor, $B_{\mu\nu}$, with curvature terms, which is given by
\begin{align*} \label{nm}
\mathcal{L}_{LVg} = \left(\tilde{\xi}_2\mathfrak{B}^\mu \mathfrak{B}^\nu +\xi_2 B^{\mu}{_\alpha}B^{\alpha\nu}\right)R_{\mu\nu}.\numberthis
\end{align*}
The Lagrangians $\mathcal{L}_{K}$ and  $\mathcal{L}_{V}$, given by
\begin{align}
\mathcal{L}_{K} =& -2\kappa\left[\frac{1}{4}\mathfrak{B}_{\mu\nu}\mathfrak{B}^{\mu\nu} + \frac{1}{12}H_{\lambda\mu\nu}H^{\lambda\mu\nu}\right],\\
\mathcal{L}_{V} =& -2\kappa\left[\mathfrak{U}(\mathfrak{B}_{\mu}\mathfrak{B}^{\mu}\pm\mathfrak{b}^2)+V(B_{\mu\nu}B^{\mu\nu}\pm b^2)\right],
\end{align}
respectively, encode the kinetic terms of the bumblebee and the antisymmetric rank-2 tensor fields and their self-interaction potentials, whilst $\mathcal{L}_{M}$ is the matter Lagrangian. Here, $\mathfrak{b}$ and $b$ are real constants, the sign in the argument of the self-interaction potentials is chosen depending on the norm of the tensor fields $\mathfrak{B}_\mu$ and $B_{\mu\nu}$, specifically plus $(+)$ if the tensor is timelike, and minus $(-)$ if it is spacelike. Moreover, $\mathfrak{B}_{\mu\nu}$ and $H_{\lambda \mu \nu}$ are the field strengths of the bumblebee field and of the rank-2 antisymmetric tensor, respectively given by
\begin{align}
\mathfrak{B}_{\mu\nu}= &\partial_\mu\mathfrak{B}_{\nu}-\partial_\nu\mathfrak{B}_{\mu},\\
H_{\lambda \mu \nu} = &\partial_{\lambda} B_{\mu \nu} + \partial_{\mu} B_{\nu \lambda} + \partial_{\nu} B_{\lambda \mu}.
\end{align}
We assume that the self-interaction potentials, $\mathfrak{U}$ and $V$, trigger spontaneous breaking of the Lorentz symmetry. They lead to the formation of non-zero VEVs for the bumblebee and the 2-form field, respectively,
\begin{align} \label{vev1}
    \langle \mathfrak{B}_{\mu} \rangle = \mathfrak{b}_{\mu},\quad\text{and} \quad \langle B_{\mu \nu} \rangle = b_{\mu \nu},
\end{align}
which define background tensor fields. Consequently, we have a local breaking of both Lorentz symmetry and diffeomorphism symmetry~\cite{Bluhm:2007bd}. In this vacuum configuration, we will explore, in the subsequent sections, the implications of Lorentz symmetry violation on the geometry of wormholes.

The VEV configuration is particularly interesting since it provides a simple framework for investigating the effects of LV in the gravitational sector. The potentials that drive the spontaneous breaking of Lorentz symmetry take a simple form in this configuration. For instance, we can consider the well-known smooth quadratic potential, extensively studied in the literature~\cite{Bluhm:2004ep,Bluhm:2007bd,Lessa:2021npz}, given by
\begin{equation}   \label{quadr}
\mathfrak{U} = \frac{1}{2} \tilde{\alpha} \mathfrak{X}^2,\,V = \frac{1}{2} \alpha X^2,
\end{equation}
where $\tilde{\alpha}$ and $\alpha$ are constants and $\mathfrak{X}\equiv \mathfrak{B}_{\mu}\mathfrak{B}^\mu\pm\mathfrak{b}^2$ and $X=B_{\mu\nu}B^{\mu\nu}\pm b^2$. To ensure that~(\ref{vev1}) represent vacuum configurations, the constant norm conditions, expressed as
\begin{align} \label{norm1}
    \mathfrak{b}^2 = &\langle \mathfrak{B}_{\mu}\mathfrak{B}^{\mu} \rangle =  g^{\mu\nu}   \mathfrak{b}_{\mu}\mathfrak{b}_{\nu},
    \end{align}
    \begin{align} \label{norm2}
b^2 = &\langle   B_{\mu\nu}B^{\mu\nu} \rangle =  g^{\alpha\mu} g^{\beta\nu}   b_{\alpha\beta}b_{\mu\nu},
\end{align}
 must be satisfied. 

As discussed in Ref.~\cite{lessa2025self}, the assumption that the background fields couple solely to the gravity sector can impose stringent constraints on the structure of the spacetime, complicating the search for compact objects in non-vacuum environments. However, by assuming the couplings of the background fields to matter may relax the constraints on the spacetime geometry induced by the LV~\cite{liu2025charged,duan2024electrically}. Here, we consider possible LV-matter couplings of the background fields to a phantom scalar field, through the matter Lagrangian $\mathcal{L}_{M}=\mathcal{L}_{\Phi}+ \mathcal{L}_{LV\Phi}$, where $\mathcal{L}_{\Phi}=\kappa[\partial_\mu \Phi\partial^\mu\Phi-2\mathcal{V}(\Phi)]$ is the standard Lagrangian of phantom scalar field with self-interaction potential $\mathcal{V}(\Phi)$, and
\begin{equation} \label{nomsc}
\mathcal{L}_{LV\Phi} = -2\kappa\left(\tilde{\eta}\mathfrak{B}_\mu\mathfrak{B}_\nu+\eta B_{\mu}{^\alpha}B_{\alpha\nu}\right)\partial^\mu\Phi\partial^\nu\Phi,
\end{equation}
encodes the LV-matter couplings, where $\eta,\tilde{\eta}$ are coupling constants.  The inclusion of these new terms not only affects the dynamics of the phantom scalar field but also modifies the gravitational dynamics. Additionally, they play a crucial role in the constraints arising from the dynamics of the background fields, as we shall see in the subsequent sections.

The gravitational field equations follow from the variation of the action (\ref{ac1}) with respect to the metric, namely
\begin{equation}
   \label{eq:einstein_KR}G_{\mu\nu} = R_{\mu\nu}-\frac{1}{2}R g_{\mu\nu} = \kappa T_{\mu\nu},
\end{equation}
where $T_{\mu\nu}= T_{\mu\nu}^{\mathfrak{B}}+T_{\mu\nu}^{B}+T_{\mu\nu}^{\Phi}+T_{\mu\nu}^{LVg} + T_{\mu\nu}^{LV \Phi}$ is the (total) energy momentum tensor, with
\begin{align*}
T_{\mu\nu}^{\mathfrak{B}} =&-\mathfrak{B}_{\mu\alpha}\mathfrak{B}^{\alpha}{}_{\nu}-\frac{1}{4}\mathfrak{B}_{\alpha\beta}\mathfrak{B}^{\alpha\beta} g_{\mu\nu} \\&- \mathfrak{U}g_{\mu\nu}+\mathfrak{U}_{\mathfrak{X}}\mathfrak{B}_{\mu}\mathfrak{B}_{\nu},\numberthis\\
T_{\mu\nu}^{B} =&\frac{1}{2}H^{\alpha\beta}{}{}_{\mu}H_{\nu\alpha\beta} - \frac{1}{12}g_{\mu\nu}H^{\alpha\beta\lambda}H_{\alpha\beta\lambda}-g_{\mu\nu}V \\&+ 4B^{\alpha}{}_{\mu}B_{\alpha\nu}V_{X},\numberthis\\
T_{\mu\nu}^{\Phi} = &-\frac{1}{2}\partial_{\mu}\Phi\partial_{\nu}\Phi+\frac{1}{4}\partial_{\alpha}\Phi \partial^{\alpha}\Phi g_{\mu\nu}-\frac{1}{2}\mathcal{V}(\Phi)g_{\mu\nu},\numberthis
\end{align*}
encoding, respectively, the minimal couplings of the bumblebee, the antisymmetric tensor and the phantom scalar field with the spacetime metric, where $\mathfrak{U_X}=\partial\mathfrak{U}/{\partial \mathfrak{X}}$ and $V_X=\partial V/{\partial X}$. Moreover,
\begin{widetext}
\begin{align*}\label{t2}
T_{\mu\nu}^{LVg} = &\frac{\tilde{\xi}_2}{\kappa}\bigg( \frac{1}{2}\mathfrak{B}^{\alpha}\mathfrak{B}^{\beta}R_{\alpha\beta}g_{\mu\nu} - \mathfrak{B}^{\alpha}R_{\alpha\mu}\mathfrak{B}_{\nu}- \mathfrak{B}^{\alpha}R_{\alpha\nu}\mathfrak{B}_{\mu}+\frac{1}{2}\nabla_{\alpha} \nabla_{\mu}(\mathfrak{B}_{\nu}\mathfrak{B}^{\alpha})+\frac{1}{2}\nabla_{\alpha} \nabla_{\nu}(\mathfrak{B}_{\mu}\mathfrak{B}^{\alpha}) \\&-\frac{1}{2}\nabla_{\alpha}\nabla^{\alpha}(\mathfrak{B}_{\mu}\mathfrak{B}_{\nu}) -\frac{1}{2}g_{\mu\nu}\nabla_{\alpha}\nabla_{\beta}(\mathfrak{B}^{\alpha} \mathfrak{B}^{\beta})\bigg) +\frac{\xi_2}{\kappa} \bigg(\frac{1}{2}B^{\alpha\lambda}B^{\beta}{}_{\lambda}R_{\alpha\beta}g_{\mu\nu} -B^{\alpha}{}_{\mu}B^{\beta}{}_{\nu}R_{\alpha\beta}\\&-B^{\alpha\beta}R_{\alpha\mu} B_{\nu\beta}-B^{\alpha\beta}R_{\alpha\nu}B_{\mu\beta}+\frac{1}{2}\nabla_{\alpha}\nabla_{\mu}(B_{\nu\beta} B^{\alpha\beta})+\frac{1}{2}\nabla_{\alpha}\nabla_{\nu}(B_{\mu\beta}B^{\alpha\beta})\\
&-\frac{1}{2}\nabla_{\lambda}\nabla^{\lambda}(B^{\alpha}{}_{\mu}B_{\alpha\nu}) -\frac{1}{2}g_{\mu\nu}\nabla_{\alpha}\nabla_{\beta}(B^{\alpha\lambda}B^{\beta}{}_{\lambda})\bigg) \numberthis
\\
T_{\mu\nu}^{LV\Phi} = &\,\tilde{\eta}\mathfrak{B}^{\alpha}\nabla_{\alpha}\Phi\left(2\mathfrak{B}_{\mu} \nabla_{\nu}\Phi +2\mathfrak{B}_{\nu}\nabla_{\mu}\Phi -g_{\mu\nu} \mathfrak{B}_{\beta}\nabla^{\beta}\Phi\right)\\&+\eta\left[ \left(2 B_{\mu\alpha}B_{\nu\beta} - B_{\alpha}{}^{\lambda}B_{\beta\lambda} g_{\mu\nu}\right)\nabla^{\alpha}\Phi\nabla^{\beta}\Phi + 2\nabla^{\alpha}\Phi B_{\alpha\beta}B_{ \nu}{}^{\beta}\nabla_{\mu}\Phi+ 2\nabla^{\alpha}\Phi B_{\alpha\beta}B_{ \mu}{}^{\beta}\nabla_{\nu}\Phi \right]\numberthis,
\end{align*}
\end{widetext}
encode the non-minimal couplings of the Lorentz-violating fields with gravity and with the scalar matter, respectively.

By varying the action~\eqref{ac1} with respect to \(\Phi\) one obtains the equations of motion for the scalar field, namely
\begin{align*}
  \label{eq:scalarfield_eq}\Box \Phi +\frac{\partial \mathcal{V}}{\partial\Phi}& = 2 \tilde{\eta} \bigg[\mathfrak{B}^{\alpha}\mathfrak{B}^{\beta}\nabla_{\alpha}\nabla_{\beta} \Phi+\nabla_\beta\Phi\nabla_{\alpha}\left(\mathfrak{B}^{\alpha} \mathfrak{B}^{\beta}\right)\bigg]\\
  &\hspace{-0.5cm} + 2 \eta \bigg[B_{\alpha}{}^{\mu}B_{\beta\mu}\nabla^{\beta} \nabla^{\alpha}\Phi+\nabla^\alpha\Phi\nabla_{\mu}\left(B^\mu{_\beta} B_\alpha{^\beta}\right)\bigg],\numberthis
 \end{align*}
where $\Box =\nabla_{\alpha}\nabla^{\alpha} $. As we can see, the LV-matter couplings of the scalar field to the bumblebee and to the antisymmetric rank-2 tensor introduce additional terms on the standard Klein-Gordon equation. The variation of the matter Lagrangian with respect to the $\mathfrak{B}_\mu$ and $B_{\mu\nu}$ yields, respectively, to
\begin{align*}
\label{eq:VEV_bumblebee}\nabla_{\mu}\mathfrak{B}^{\mu\nu} =&2 \mathfrak{U}_{\mathfrak{X}}\mathfrak{B}^{\nu} - \frac{\tilde{\xi}_2}{\kappa}\mathfrak{B^{\mu}}R_{\mu}{}^{\nu} +2 \tilde{\eta} \mathfrak{B}^{\mu}\nabla_{\mu}\Phi\nabla^{\nu}\Phi,\numberthis\\
\label{eq:VEV_H}\nabla_{\alpha}H^{\alpha\mu\nu} = &4 V_X B^{\mu\nu}  +\frac{\xi_2}{\kappa}B_{\alpha}{}^{\mu}R^{\nu\alpha} -\frac{\xi_2}{\kappa}B_{\alpha}{}^{\nu}R^{\mu\alpha} \\& +2 \eta B^{\mu}{}_{\alpha}\nabla^{\nu}\Phi \nabla^{\alpha}\Phi-2 \eta B^{\nu}{}_{\alpha}\nabla^{\mu}\Phi \nabla^{\alpha}\Phi\numberthis.
\end{align*}
Therefore, in order to search for wormhole solutions supported by phantom fields with LV, we must solve the metric equation~\eqref{eq:einstein_KR} along with the dynamical equations for the phantom field~\eqref{eq:scalarfield_eq} and the dynamical equations for the tensors with non-zero VEV, namely Eqs.~\eqref{eq:VEV_bumblebee} and~\eqref{eq:VEV_H}. As we shall see, for suitable VEV configurations, these latter equations act as constraints on the spacetime geometry and on the scalar field profile.


\section{Wormholes supported by phantom distributions coupled to Lorentz-violating fields} \label{seciv}

In this and in the forthcoming sections, we explore the implications of allowing LV-matter distributions in the spacetime. A remarkable consequence of the inclusion of these additional terms is the possibility of simple scalar field distributions acting as sources of wormholes in Lorentz-violating backgrounds. We remark that phantom wormholes in gravity frameworks considering LV-gravity couplings were already found, for instance, in Ref.~\cite{magalhaes2025wormholes}, where this wormhole-type was found by considering self-interacting phantom scalar fields in a Lorentz-violating model where the Riemann tensor couples with an antisymmetric rank-2 tensor with non-vanishing VEV. Surprisingly, one of the phantom wormholes found is a simple extension of the (symmetric) EBWH, namely
\begin{equation}
    \label{eq:EBHW_LV}ds^2=-dt^2+dx^2+\left(a^2+\dfrac{x^2}{1-L}\right)\left(d\theta^2+\sin^2\theta d\varphi^2\right),
\end{equation}
where $L$ encodes the contributions of the VEV of the antisymmetric tensor $B^{\mu\nu}$ and its non-minimal coupling to the Riemann tensor. As discussed in Ref.~\cite{lessa2025self}, phantom wormholes of this type are absent in LV backgrounds when considering non-minimal couplings to the Ricci tensor. This absence arises due to the anisotropies introduced by the non-vanishing VEV fields. However, in this work, we show that the inclusion of additional LV-matter couplings enables the emergence of such wormhole solutions in LV models with non-minimal couplings to the Ricci tensor. Furthermore, these solutions are supported by free scalar fields, providing a new perspective on the viability of phantom wormholes in LV frameworks.

Let us consider static and spherically symmetric wormholes with line element given by
\begin{equation} \label{metric}
   ds^2 = -A(x)dt^2 + \dfrac{dx^2}{A(x)} + r(x)^2 d\Omega^2,
\end{equation}
where the radial coordinate $x\in (-\infty,\infty)$ and $d\Omega^2=d\theta^2+\sin^2\theta d\varphi^2$ is the line element of a unit sphere. Since we are looking for wormhole solutions, the areal radius $r(x)$ is assumed to possess a minimum, that without loss of generality can be taken at $x=0$, thus
\begin{equation} \label{areal}
r(0)=a, \quad    r'(0)=0, \quad  r''(0)>0
\end{equation}
where $a$ is the radius of the wormhole throat. Moreover, the primes denotes derivatives with respect to the radial coordinate $x$.

Since we are adopting the line element~\eqref{metric}, a stationary, spacelike VEV of the bumblebee field given by $\mathfrak{b}_{\mu}=(0,\mathfrak{b} /\sqrt{A(x)},0,0)$ ensures a constant norm, $\mathfrak{b}_{\mu} \mathfrak{b}^{\mu} =\mathfrak{b}^2$. While for the antisymmetric rank-2 tensor, as proposed in Ref.~\cite{Lessa:2019bgi}, it is convenient to express the VEV as $b_{\mu \nu} = \tilde{E}_{[\mu} v_{\nu]} + \epsilon_{\mu \nu \alpha \beta} v^{\alpha} \tilde{B}^{\beta}$, where the background vectors \( \tilde{E}_{\mu} \) and \( \tilde{B}_{\mu} \) can be interpreted as pseudo-electric and pseudo-magnetic fields, respectively, and \( v^{\mu} \) is a timelike 4-vector. The pseudo-fields \( \tilde{E}_{\mu} \) and \( \tilde{B}_{\mu} \) are spacelike, i.e., $\tilde{E}_{\mu} v^{\mu} = \tilde{B}_{\mu} v^{\mu} = 0.$ Here we consider a pseudo-electric configuration of 2-form $\textbf{b}_{2} = - \tilde{E}(x) \ dt \wedge dx$, i.e., the only nonvanishing terms are $b_{tx}=-b_{xt}= - \tilde{E}(x)$, which can be determined through Eq.~\eqref{norm2} for the line element~\eqref{metric}, yielding a constant pseudo-electric field, expressed as $\tilde{E} = b/\sqrt{2}$. As a consequence, both VEV's field strengths vanish, namely $\langle \mathfrak{B}_{\mu\nu} \rangle=0$ and $\langle H_{\mu\nu\alpha}\rangle=0$.

A direct implication of the vanishing of $\langle \mathfrak{B}_{\mu\nu}\rangle$ and $\langle H_{\mu\nu\alpha}\rangle$ is that the dynamical equations~\eqref{eq:VEV_bumblebee} and~\eqref{eq:VEV_H} turn into constraints on the geometry and on the matter distributions. In the VEV configuration, the anisotropies induced by the LV are encoded in
the VEV equation, which, for the bumblebee and for the antisymmetric rank-2 tensor, are, respectively,
\begin{align*}
& \label{eombum}-\frac{\tilde{\xi}_2}{\kappa}\mathfrak{b^{\mu}}R_{\mu}{}^{\nu} +2 \tilde{\eta} \mathfrak{b}^{\mu}\nabla_{\mu}\Phi\nabla^{\nu}\Phi=0,\numberthis \\
&\label{eomkr}\frac{\xi_2}{\kappa}b_{\alpha}{}^{\mu}R^{\nu\alpha} -\frac{\xi_2}{\kappa}b_{\alpha}{}^{\nu}R^{\mu\alpha}  +2 \eta b^{\mu}{}_{\alpha}\nabla^{\nu}\Phi \nabla^{\alpha}\Phi\\&-2 \eta b^{\nu}{}_{\alpha}\nabla^{\mu}\Phi \nabla^{\alpha}\Phi=0,\numberthis
\end{align*}
where we have used that $\langle\mathfrak{U}_\mathfrak{X}\rangle=\langle V_X\rangle=0$, since we are considering a quadratic self-interaction potential~\eqref{quadr}. One notices that the LV-matter couplings between the background fields and the derivatives of the scalar field modify the constraints imposed by the VEV equations. In the absence of these additional couplings, as discussed in Ref.~\cite{lessa2025self}, tideless wormholes---characterized by a lapse function $A=1$---such as the one described by the line element~\eqref{eq:EBHW_LV}, cannot emerge in such LV gravity frameworks. However, these wormholes can arise in LV scenarios if suitable coupling constants are considered. To explore tideless wormholes within the gravity framework given by Eq.~\eqref{ac1}, we will henceforth assume $A=1$.  It can be observed that by substituting Eq.~\eqref{metric} into Eqs.~\eqref{norm1} and~\eqref{norm2} yields,  respectively,
\begin{equation}
\label{eq:VEV_config}\mathfrak{b}_{\mu}=\begin{pmatrix}
0,\mathfrak{b},0,0
\end{pmatrix},\quad    b_{\mu\nu}=\begin{pmatrix}
0 & -\frac{b}{\sqrt{2}} & 0 & 0 \\
\frac{b}{\sqrt{2}} & 0 & 0 & 0 \\
0 & 0 & 0 & 0 \\
0 & 0 & 0 & 0
\end{pmatrix}.
\end{equation}
Therefore, the VEV equation of the bumblebee~\eqref{eombum} and of the antisymmetric rank-2 tensor field~\eqref{eomkr} become, respectively,
\begin{align}
\label{eq:VEV_V_constraint2}
\mathfrak{b}\bigg(\frac{\tilde{\xi}_2r''}{\kappa r} + \tilde{\eta}\Phi'^2\bigg) = 0.\\
\label{eq:VEV_V_constraint}
b\bigg(\frac{\xi_2r''}{\kappa r} + \eta\Phi'^2\bigg) = 0.
\end{align}
At this stage, the necessity of introducing the LV-matter coupling becomes evident: for $\eta=\tilde{\eta}=0$, Eqs.~\eqref{eq:VEV_V_constraint2} and~\eqref{eq:VEV_V_constraint} are purely geometric constraints that eliminate the possibility of a local minimum in the areal function, thus ruling out the existence of a wormhole solution.

Under the VEV configuration~\eqref{eq:VEV_config}, the gravitational field equations~\eqref{eq:einstein_KR} reduce to the three differential equations:
\begin{widetext}
\begin{align}
\label{eq_system_1}&\frac{2 r r''+r'^2-1}{r^2}=-\frac{1}{2} \kappa   \left(2 \tilde{\eta}\mathfrak{b}^2+\eta b^2 -1\right)\Phi'^2 + \frac{ \left(b^2 \xi_2-2 \mathfrak{b}^2 \tilde{\xi}_2\right)r'^2-4 \mathfrak{b}^2 \tilde{\xi}_2 r r''}{2 r^2},\\
\label{eq_system_2}&\frac{r'^2-1}{r^2}=\frac{1}{2} \kappa  \left(6\tilde{\eta} \mathfrak{b}^2 -3 \eta b^2 -1\right)\Phi'^2 -\frac{\left(2 \mathfrak{b}^2 \tilde{\xi}_2-b^2\xi_2\right)\left(r'^2-2 r r''\right)}{2 r^2},\\
\label{eq_system_3}&\frac{r''}{r}=-\frac{1}{2} \kappa   \left(2 \tilde{\eta} \mathfrak{b}^2 -\eta b^2 -1\right)\Phi'^2+ \frac{r'' \left(b^2\xi_2-2 \mathfrak{b}^2 \tilde{\xi}_2\right)}{2 r},
\end{align}
\end{widetext}
while, the scalar field equation~\eqref{eq:scalarfield_eq} reads
\begin{equation} \label{phieom}
  (-1+2\tilde{\eta}\mathfrak{b}^2-\eta b^2) \bigg( \Phi''+\frac{2r'\Phi'}{r} \bigg)  = 0,
\end{equation}
where we recall that we are considering a free scalar field, such that $\partial\mathcal{V}/\partial \Phi=0$, and we are assuming that the scalar field inherits the symmetries of the spacetime, thus it depends solely on the radial coordinate, namely $\Phi\equiv\Phi(x)$.

By subtracting Eqs.~\eqref{eq_system_1} and~\eqref{eq_system_2}, we obtain an expression for the derivative of the scalar field with respect to the radial coordinate $x$, given by:
\begin{equation} \label{phimixed}
    \Phi '^2= \frac{2  \left(l_T-2 l_V-1\right)r''}{ \left(4 \mathfrak{b}^2 \tilde{\eta}-b^2 \eta-1\right)\kappa  r},
\end{equation}
where we have introduced $l_V=\tilde{\xi}_2\mathfrak{b}^2$ and $l_T=\xi_2 b^2/{2}$. Substituting the above equation into the VEV constraints~\eqref{eq:VEV_V_constraint2} and~\eqref{eq:VEV_V_constraint}, we obtain the following algebraic equations
\begin{align}
    &2 (l_T-1)\tilde{\eta}-\frac{2 l_T \tilde{\xi}_2}{\xi_2}\eta-\tilde{\xi}_2=0,\\
   &\frac{4  l_V \xi_2}{\tilde{\xi}_2}\tilde{\eta}-2  (2 l_V+1)\eta-\xi_2=0.
\end{align}
It is straightforward to verify that, these constraints are satisfied only if
\begin{align}
\label{xi1}\tilde{\eta}&= -\frac{\tilde{\xi}_2}{2},\\
\label{etaxi}\eta &= -\frac{\xi_2}{2}.
\end{align}

Similarly, we can subtract Eq.~\eqref{eq_system_1} from Eq.~\eqref{eq_system_3}, and then substitute Eq.~\eqref{phimixed} in it. Then, by applying the relations~\eqref{xi1} and~\eqref{etaxi}, we find a differential equation solely for the areal radius, namely
\begin{equation}\label{arealmixed}
    r  \left(-l_T+l_V+1\right)r''+ \left(-l_T+l_V+1\right)r'^2-1=0.
\end{equation}
By imposing that there is a throat structure at $x=0$, we obtain
\begin{equation}
\label{eq:r_solution_ellismixed}r(x)=\sqrt{a^2+\frac{x^2}{1+l_V-l_T}},
\end{equation}
We note that, in order to ensure that the above areal radius presents a local minimum, we require that $l_{V}-l_{T}>-1$.

The phantom field that supports such structure is obtained by substituting the areal radius~\eqref{eq:r_solution_ellismixed} into~\eqref{phimixed}, with~\eqref{xi1} and~\eqref{etaxi}, and integrating the resulting equation, where we find that
\begin{equation}
  \label{eq:ellis_wormholemixedd}  \Phi = \sqrt{2/\kappa}\arctan\left(\frac{x/a}{\sqrt{1+l_V-l_T}}\right),
\end{equation}
where we have chosen, without loss of generality, the solution with plus sign and with vanishing integration constant.

The resulting spacetime described by the metric
\begin{equation}
  \label{eq:ellis_wormholemixed}  ds^2 = -dt^2+dx^2+\left(a^2+\frac{x^2}{1+l_V-l_T}\right)d\Omega^2,
\end{equation}
corresponds to an EB-like wormhole, which is supported by a phantom scalar field coupled to both the bumblebee field and the antisymmetric tensor field, each possessing a non-zero VEV. Hereafter, one refers to such a spacetime as a Lorentz-violating Ellis-Bronnikov (LVEB) wormhole. If  simultaneously $l_V\to 0$ and $l_T\to 0$, we recover the EB wormhole solution. Notably, if the LV parameters satisfy $l_V=l_T$, the line element in  Eq.~\eqref{eq:ellis_wormholemixed} also simplifies to that of the EB wormhole. This result puts light on the importance of verifying whether more astrophysically relevant solutions, such as black holes, can incorporate distinct combinations of LV parameters. The outcome also indicates that experiments focusing solely on quantities minimally coupled to the spacetime metric might miss LV signatures when different Lorentz-violating fields are involved.

\begin{figure}[!htb]
    \centering
    \includegraphics[width=\columnwidth]{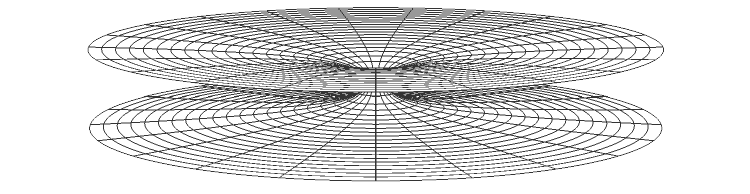}
\includegraphics[width=\columnwidth]{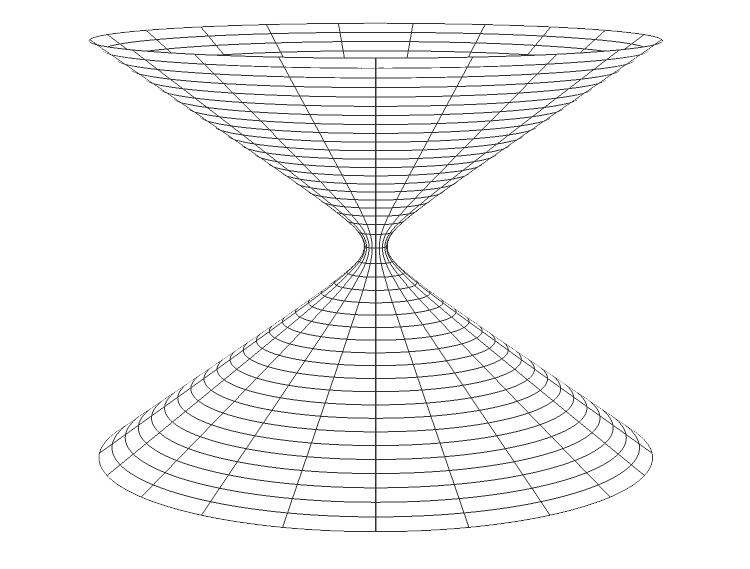}
    \caption{Embedding diagrams of the equatorial plane of EB wormhole (top) and LVEB wormholes with $l_V-l_T=0.5$ (bottom).}
    \label{fig:embed}
\end{figure}

Before we end this section, let us discuss some distinct characteristics that LVEB wormholes have compared to standards EB wormholes. Notably, the wormhole reported here has the same qualitative structure as wormhole~\eqref{eq:EBHW_LV}. Several notable properties of this spacetime have been investigated in Ref.~\cite{magalhaes2025wormholes}. For instance, far from the throat these wormholes does not approach a flat region, rather their equatorial plane display a conical geometry with a deficit (or excess) angle $\delta$, specifically it reads $\delta = \pi(l_T-l_V)/(l_T-l_V-1)$. Such structure can be visualized in Fig.~\ref{fig:embed}, where we exhibit the embedding diagrams of the equatorial plane of a LVEB wormhole with $l_T-l_V=0.5$. Due to this conical behavior, the geodesic structure of the LVEB wormholes is clearly told apart from the one of EB wormholes. As a consequence, the lensing properties of these wormholes change compared to EB wormholes, even in the weak-field approximation, where one finds that the total deflection angle of massless particles bears deviations induced by the LV. Remarkably, configurations with $l_V=l_T$ do not exhibit any angle defect, i.e., $\delta=0$, once the EB geometry is restored.

It is worth mentioning that, similar to EB wormholes, these tideless wormholes possess a single light ring located at their throats. One can analyze the (in)stability of these light rings using the principal Lyapunov exponents. Specifically, for massless particles in circular motion at the throat of an LVEB wormhole, the principal Lyapunov exponent reads~\cite{magalhaes2024echoes}
\begin{equation}
\lambda =  \sqrt{\frac{r''(0)}{r(0)}} = \dfrac{1}{a\sqrt{1+l_V-l_T}}.
\end{equation}
Thus, the light rings at the throat correspond to unstable circular orbits, where one recalls that the inequality $1+l_V-l_T>0$ must be satisfied for the throat-like structure to be present. Remarkably, the LV influences the instability of circular motion at the throat. Nonetheless, we point out that light rings at the throats of configurations with $l_V=l_T$ share the same Lyapunov exponents as the ones in EB wormholes.

\section{Dynamics of scalar perturbations in Lorentz-violating wormholes}\label{sec:dynamics}
The study of how the LVEB wormhole models, introduced in the previous section, respond to perturbations provides critical insights into how the breaking of Lorentz symmetry affects the structure and stability of wormholes. To gain valuable intuition into these effects, in this and the following sections, we will analyze the response of the LVEB wormhole models to a massless (canonical) test scalar field, denoted by $\Psi$. This scalar field serves as the simplest wave-like probe that one can introduce into the geometry. It is important to note that the scalar field $\Psi$ is distinct from the phantom scalar field $\Phi$, which is responsible for generating the wormhole geometry. Furthermore, we assume that the scalar field $\Psi$ does not backreact on the underlying spacetime.

The dynamics of a canonical test scalar field in a LVEB wormhole background depends critically on whether the field is assumed to solely minimally couple to the metric tensor or if it also couples with the Lorentz-violating fields. This distinction plays a crucial role in determining the behavior and evolution of the scalar field within the LVEB wormhole framework. Let us consider both scenarios and highlight their differences. In order to facilitate the study, let us split the analysis and denote by $\Psi_{g}$ a canonical scalar field minimally coupled to the spacetime metric and by $\Psi_{LV}$ a canonical scalar field coupled to both the spacetime metric and to Lorentz-violating fields. 

Hence, the dynamics of $\Psi_{g}$ is described by the standard Klein-Gordon equation, namely
\begin{equation}
\label{eq:KG_pert}
    \Box\Psi_g=0,
\end{equation}
while the dynamics of $\Psi_{LV}$ is described by, cf. Eq.~\eqref{eq:scalarfield_eq},
\begin{align}
\label{eq:scalarfield_eq_sintetica}\nabla_\mu(q^{\mu\nu}\nabla_\nu\Psi_{LV})=0,
 \end{align}
where, under the VEV configurations considered here, $q^{\mu\nu}:= g^{\mu\nu}+2\tilde{\zeta}\mathfrak{b}^\mu\mathfrak{b}^\nu+2\zeta b^{\mu\alpha}b^\nu{_\alpha}$, where $\tilde{\zeta}$ and $\zeta$ are coupling constants, and we have used the metric compatibility condition. We remark that, since $\Psi_{LV}$ is not the phantom scalar field that supports the LVEB wormholes, it does not necessarily respond to the LV-matter coupling in the same way as $\Phi$. Specifically, the LV-matter coupling constants $\tilde{\zeta}$ and $\zeta$ are not the same as $\tilde{\eta}$ and $\eta$. Furthermore, while $\tilde{\eta}$ and $\eta$ are constrained by Eqs.~\eqref{eq:VEV_V_constraint2} and~\eqref{eq:VEV_V_constraint}, the constants $\tilde{\zeta}$ and $\zeta$ are, in principle, unconstrained.

Let us consider that the scalar perturbations can be decomposed using the spherical harmonics, $Y_{\ell m}(\theta,\varphi)$, namely
\begin{align}
\label{eq:scalarfield_decomp1}\Psi_g &= \sum_{\ell,m}\int d\omega\frac{ \psi_g(x)}{r(x)}e^{-i\omega t}Y_{\ell m}(\theta,\varphi),\\\Psi_{LV} &= \sum_{\ell,m}\int d\omega\frac{ \psi_{LV}(x)}{r(x)}e^{-i\omega t}Y_{\ell m}(\theta,\varphi),\label{eq:scalarfield_decomp2}
\end{align}
where $Y_{\ell m}(\theta,\varphi)  \propto P_{\ell m}(\cos \theta) e^{im \varphi}$, with $P_{\ell m}(\cos \theta )$ being the associated Legendre polynomials, $\omega$ is the frequency, $\ell$ and $m$ are, respectively, the polar and azimuthal indices and $\psi_g(x)$ and $\psi_{LV}(x)$ are the radial parts of the perturbations $\Psi_g$ and $\Psi_{LV}$, respectively.
Due to the spherical symmetry, henceforth, $m$ is set to zero. By plugging the decompositions \eqref{eq:scalarfield_decomp1} and \eqref{eq:scalarfield_decomp2} into the scalar field equations~\eqref{eq:KG_pert} and~\eqref{eq:scalarfield_eq_sintetica}, correspondingly, one obtains that the radial part of the scalar perturbations $\Psi_g$ and $\Psi_{LV}$ are determined, respectively, by the Schr\"odinger-like equations
\begin{align}
\label{eq:schrodinger-like}
\psi_g'' =  U^g_{\omega\ell}(x)\psi_g,\\
\label{eq:schrodinger-like_II}\psi_{LV}'' =   U^{LV}_{\omega\ell}(x)\psi_{LV}
\end{align}
where $U^{g}_{\omega\ell}(x)$ and $U^{LV}_{\omega\ell}(x)$ are the effective potentials the scalar perturbations $\Psi_g$ and $\Psi_{LV}$ are subjected to, respectively.

Explicitly, the radial part of the scalar perturbations, $\psi_g$ and $\psi_{LV}$, satisfy, respectively,
\begin{widetext}
\begin{align}
\label{eq:caseIII_phatom_pert}
&\psi_g''+\left[\omega^2-\frac{(1+l_V-l_T)(1+\ell(\ell+1)(1+l_V-l_T))a^2+\ell(\ell+1)x^2}{\left(a^2(1+l_V-l_T)+x^2\right)^2}\right]\psi_g(x)=0,\\
\label{eq:caseIII_phatom_pert_2_complete}
&\frac{d^2\psi_{LV}}{d\varrho^2} + \left[\omega^{2}- \frac{(1+l_V-l_T)[a^2(1-m_T)(\ell(\ell+1)(1+l_V-l_T)+1+m_V-m_T)+\ell(\ell+1)F\varrho^2]}{(1-m_T)\left[a^2(1+l_V-l_T)+F\varrho^2\right]^2}\right]\psi_{LV}(\varrho)=0,
\end{align}
\end{widetext}
where, in Eq.~\eqref{eq:caseIII_phatom_pert_2_complete}, we have introduced a rescaled radius $\varrho=x\sqrt{1/F}$, with $F=\tfrac{1+m_V-m_T}{(1-m_T)}$. Besides, we have also adopted the LV-matter parameters $m_V=2\tilde{\zeta}\mathfrak{b}^2$ and $m_T=\zeta b^2$. As one can see, the LV-matter couplings can sharply influence the dynamics of perturbations $\Psi_{LV}$ and significantly depart it from the dynamics of $\Psi_g$. However, as one expects, by simultaneously vanishing the parameters $m_V$ and $m_T$,  the dynamics of $\Psi_{LV}$ and $\Psi_g$ become the same.

Since $\tilde{\zeta}$ and $\zeta$ are unconstrained, the LV-matter parameters are likewise unconstrained, such that the influence of the Lorentz violation in the matter sector, controlled by $m_V$ and $m_T$, can contribute differently from the gravitational sector, controlled by $l_V$ and $l_T$. Here, let us restrict our analysis to a particular yet intriguing case in which the influence in both sectors are the same, namely $m_V=l_V$ and $m_T=l_T$. In this scenario, the effective potential of $\Psi_{LV}$ reduces to
\begin{equation}
\label{eq:caseIII_phatom_pert_2}U^{LV}_{\omega\ell}(\varrho)=-\omega^{2}+ \frac{-a^2 (l_T-1) \left(\ell^2+\ell+1\right)+\ell (\ell+1) \varrho^2}{\left(\varrho^2-a^2 (l_T-1)\right)^2}.
\end{equation}
At first glance, this result may appear counterintuitive: it suggests that if $l_T$ vanishes, the scalar field perturbation $\Psi_{LV}$ is unaffected by the LV, since its differential equation becomes identical to that of the radial part of a scalar perturbation in symmetric (massless) EBWHs~\cite{blazquez-salcedoScalarAxialQuasinormal2018}. Conversely, the dynamic of $\Psi_g$ is influenced by the effects of both the non-vanishing VEVs of the antisymmetric tensor and the bumblebee fields. The fact that $\Psi_{LV}$ can propagate as if it were in an effective GR background is what one will refer to as the ``masking'' effect, and it is a direct consequence of the ``sector equality'' condition $m_V=l_V$ and $m_T=l_T$. In Sec.~\ref{sec:a_note}, we discuss the conditions under which such an effect occurs in different Lorentz-violating backgrounds.

\begin{figure}
    \centering
    \includegraphics[width=\columnwidth]{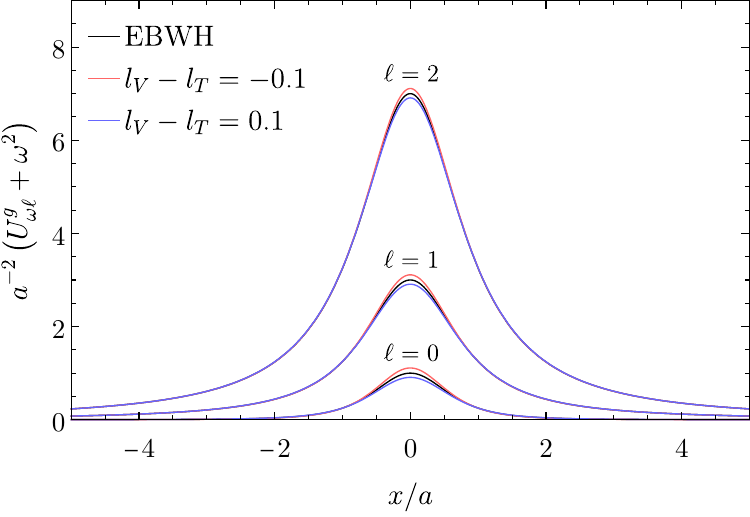}\hspace{0.1cm}\includegraphics[width=\columnwidth]{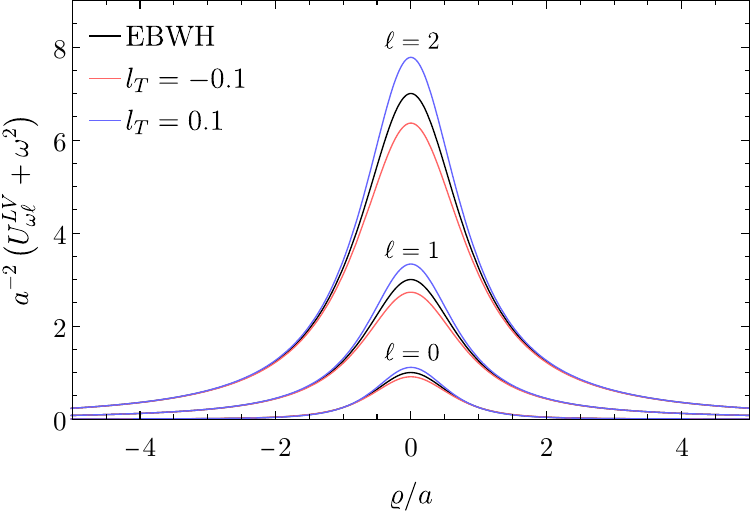}
    \caption{Frequency-independent part of the effective potentials $U_{\omega\ell}^{g}$ (top panel) and $U_{\omega\ell}^{LV}$ (bottom panel) as a function of the radial coordinates $x$ and $\varrho$, respectively. The black line corresponds to effective potential of a scalar field lying in a EBWH spacetime.}
    \label{fig:EffectivePot}
\end{figure}

\subsection{Effective potential}

Part of understanding the behavior of perturbations in the vicinity of compact objects relies on understanding the behavior of the effective potential to which the perturbation is subjected. As we have seen, if the scalar perturbation under study is assumed to be solely minimally coupled to the metric, the contributions of the spacelike VEV configuration $\mathfrak{b}^\mu$ and on the timelike VEV configuration $b^{\mu\nu}$ are explicit in the effective potential. However, if the scalar perturbation under study is assumed to be subjected to additional LV-matter couplings, the contributions of the spacelike VEV configuration $\mathfrak{b}^\mu$ are suppressed by a rescaling of the radial coordinate. Conversely, the same is not true for the timelike VEV configuration $b^{\mu\nu}$, where no rescaling of the radial coordinate removes the contribution of $l_T$ from the radial part of the perturbation.

First, let us analyze the behavior of $U_{\omega\ell}^{g}$, the effective potential that $\Psi_g$ is subjected to, for the LVEB wormhole models, given by 
\begin{align}
\label{eq:Ug}    U_{\omega\ell}^{g}=&-\omega^2+\frac{(1+l_V-l_T)(1+\ell(\ell+1)(1+l_V-l_T))a^2}{ \left(a^2(1+l_V-l_T)+x^2\right)^2} \nonumber\\
    &+\frac{\ell(\ell+1)x^2}{\left(a^2(1+l_V-l_T)+x^2\right)^2}.
\end{align}
Interestingly, if $l_V=l_T$, the effective potential becomes that of scalar perturbations in EBWHs, regardless of the absolute values of the LV parameters. One can check that the asymptotic behavior of the effective potential is
\begin{equation}
    \lim_{x\to\pm\infty}U_{\omega\ell}^g=-\omega^2,
\end{equation}
That is, it approaches negative constant values at both infinities, regardless of the values of $l_V$ and $l_T$. One can check that for $l_T-l_V<1/2$, regardless of the values of $\ell$, the effective potential has only one critical point located at the throat. For $\ell\neq 0$, the effective potential can present more than one critical point. Specifically, for $\ell=1$, the effective potential has three critical points if $l_T-l_V=1$, and for $\ell\geq2$, the effective potential has three critical points if $1/2+1/\ell(\ell+1)<l_T-l_V\leq 1$. Since we are considering solely small deviations introduced from the LV, the typical profile of the effective potential shall exhibit only one critical point, specifically a maximum. We show in the left panel of Fig.~\ref{fig:EffectivePot} the radial dependence of the effective potential, that is, $U_{\omega\ell}^g+\omega^2$. Moreover, one can see that as $l_V-l_T$ diminishes, the peak of the effective potential gets higher.


Recalling we are considering the case where the Lorentz violation influences equally the gravitational and matter sectors, $U_{\omega\ell}^{LV}$ is given by Eq.~\eqref{eq:caseIII_phatom_pert_2} and one notices that the dependence of the effective potential on the frequency decouples from the dependence on the radial coordinate. One can check that, the asymptotic behavior of the effective potential is
\begin{equation}
    \lim_{\varrho\to\pm\infty}U_{\omega\ell}^{LV}=-\omega^2,
\end{equation}
hence, it also converges to negative constant values at both infinities. Additionally, for $l_T<1$, the effective potential has a single peak located at $x=0$. In the right panel of Fig.~\ref{fig:EffectivePot}, we plot the profile of the part of the effective potential that does not depend on $\omega$, that is, $U_{\omega\ell}^{LV}+\omega^2$. Furthermore, we see that as $l_T$ increases, the peak of the effective potential increases.

\section{Quasinormal frequencies}\label{sec:qnm}

QNMs represent the characteristic oscillations of a perturbed physical system. In the context of wormholes, the boundary conditions imposed on the perturbations require purely outgoing waves at both spatial infinities, expressed as
\begin{equation}
    \label{eq:BC}\lim_{x \to \pm \infty} \psi \sim \exp(\pm i \omega x).
\end{equation}
Each mode is characterized by a complex frequency, $\omega$, where the real part, $\text{Re}[\omega]$, corresponds to the standard oscillation frequency, and the imaginary part, $\text{Im}[\omega]$, determines the rate of damping or amplification of the wave. Specifically, a negative $\text{Im}[\omega]$ indicates damping, while a positive $\text{Im}[\omega]$ signifies amplification. These modes play a crucial role in analyzing the ringdown phase following the coalescence of compact objects~\cite{kokkotas1999quasi, berti2009quasinormal}, such as black holes or neutron stars. Additionally, QNMs provide valuable insights into the stability (or instability) of such compact objects~\cite{blazquez2018scalar}, making them a key tool in gravitational wave astronomy and theoretical physics.

The study of the QNMs of EBWHs was extensively performed in the literature~\cite{konoplya2016wormholes,konoplya2018tell,blazquez2018scalar,bronnikov2021general}. In this section, we provide the analysis of the quasinormal frequencies associated to test scalar fields lying in the LVEB-wormhole backgrounds. As we have seen, the assumption of LV-matter couplings yields non-trivial contributions to the scalar dynamics; thus, one expects that the QNM profiles of the scalar perturbations $\Psi_g$ and $\Psi_{LV}$ are different.
\subsection*{Results}
We present a selection of our numerical results for the spectrum of QNMs of scalar field perturbations within a Lorentz-violating background. These results were obtained using the methods discussed in the Appendix~\ref{app:methods}, considering both scalar perturbations minimally coupled to the geometry and scalar perturbations with additional LV-matter couplings. Here, we will focus on the fundamental modes associated with $n=0$.

\subsubsection{Canonical scalar field minimally coupled to the spacetime metric}

\begin{figure*}[!htb]
    \centering
    \includegraphics[width=\columnwidth]{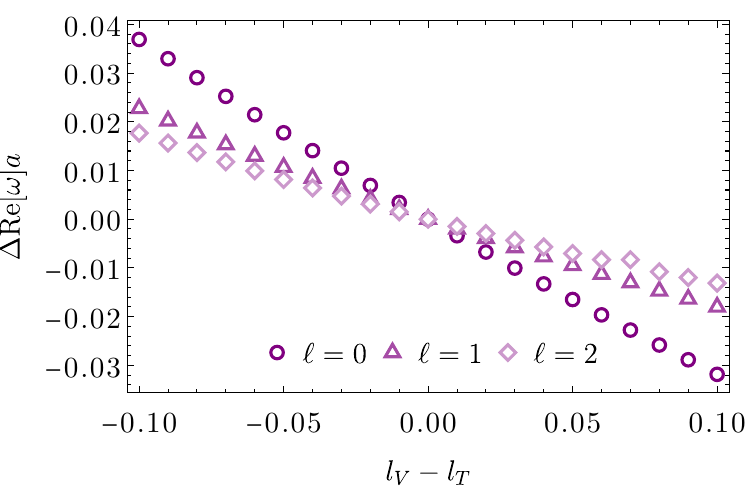} \includegraphics[width=\columnwidth]{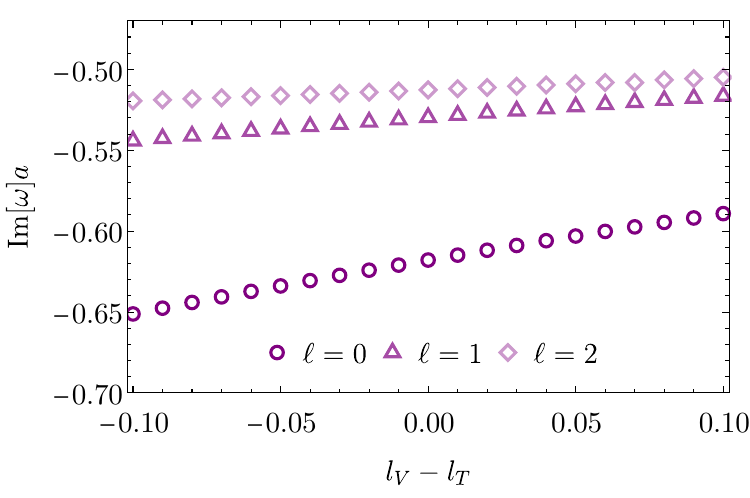}
    \caption{QNMs of scalar fields minimally coupled to gravity in LVEB wormholes, computed via DI. Left panel: Difference between the real part of the QNMs in LVEB wormholes and the corresponding real part in EBWHs. For $\ell=0$, $1$, and $2$, $\text{Re}[\omega_0]a=0.68143$, $1.5727$, and $2.5466$, respectively. Right panel: Imaginary part of the QNMs.}
    \label{fig:QNM_Re_and_Im_g}
\end{figure*}

First, let us consider the results for the scalar perturbation $\Psi_g$. In Fig.~\ref{fig:QNM_Re_and_Im_g} we show the real and the imaginary parts of the QNM frequencies, computed using DI, of scalar perturbations minimally coupled to gravity within LVEB wormholes. In this scenario, the value of $\text{Re}[\omega]$ decreases as the difference between LV parameters $l_V-l_T$ increases. In order to better appreciate this behavior, we show in Fig.~\ref{fig:QNM_Re_and_Im_g} the difference $\Delta \text{Re}[\omega] = \text{Re}[\omega] - \text{Re}[\omega_0]$, where $\omega_0$ denotes the corresponding QNM of the EBWH. Conversely, $\text{Im}[\omega]$ increases as $l_V-l_T$ increases, indicating that the modes are less damped as we increase the LV parameter. We observe that the QNMs of these scalar fields in the LVEB wormhole are always damped, indicating stable perturbations.

\begin{figure*}[!htb]
    \centering
    \includegraphics[width=\columnwidth]{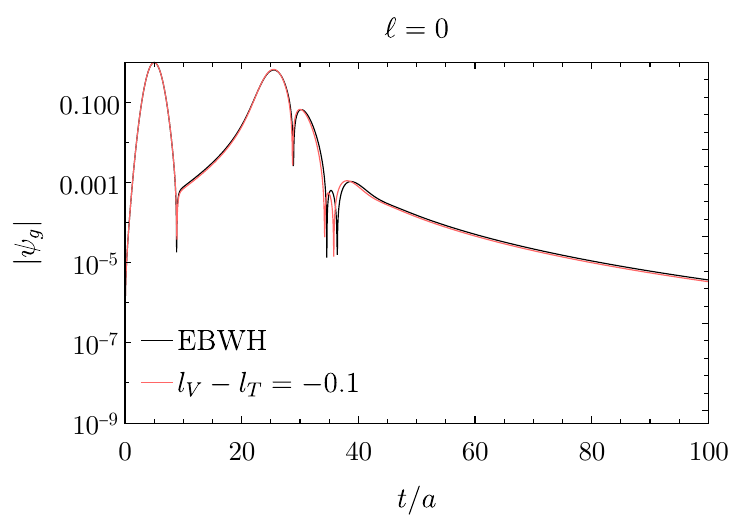}\includegraphics[width=\columnwidth]{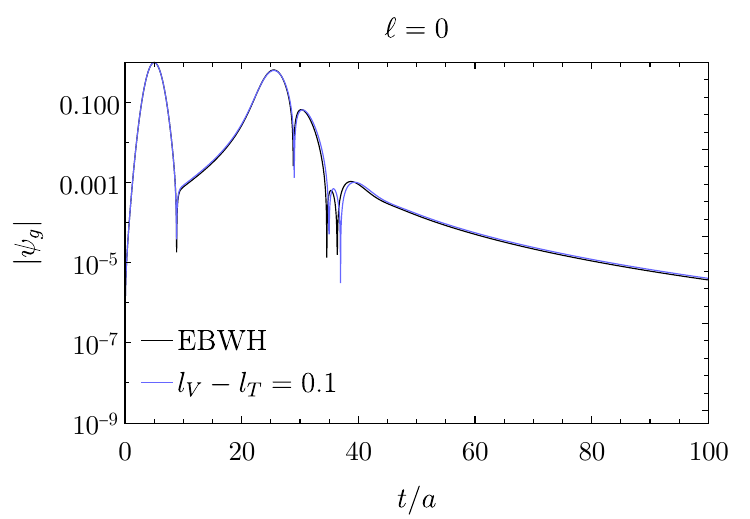}\\
    \includegraphics[width=\columnwidth]{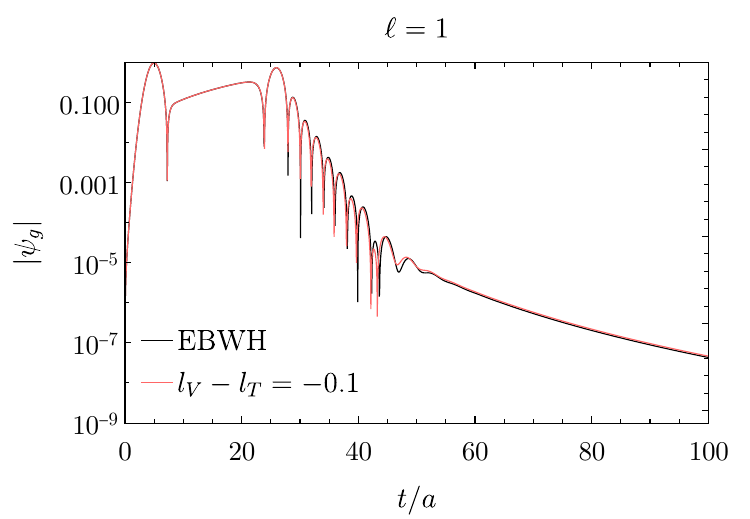}\includegraphics[width=\columnwidth]{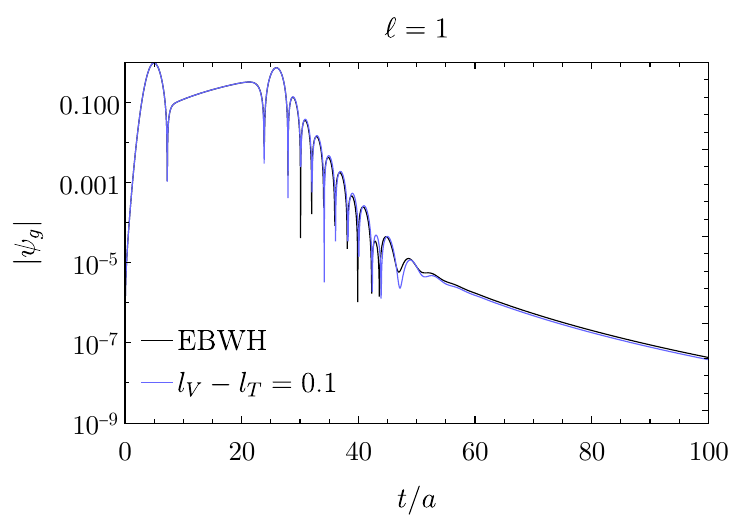}\\
    \includegraphics[width=\columnwidth]{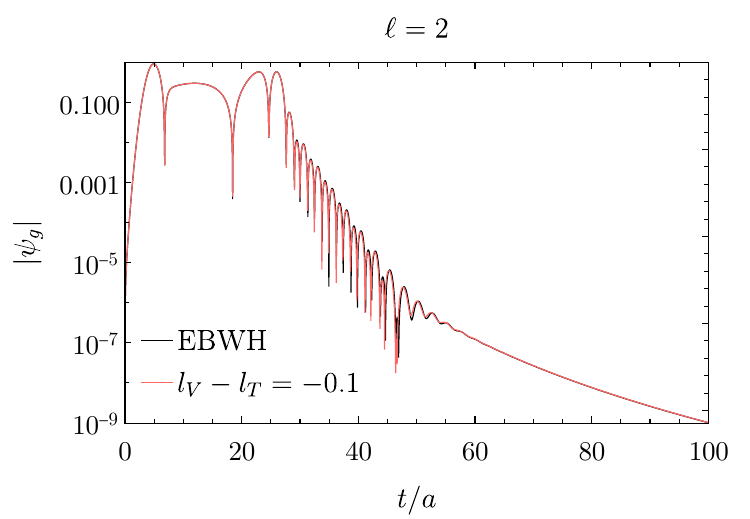}\includegraphics[width=\columnwidth]{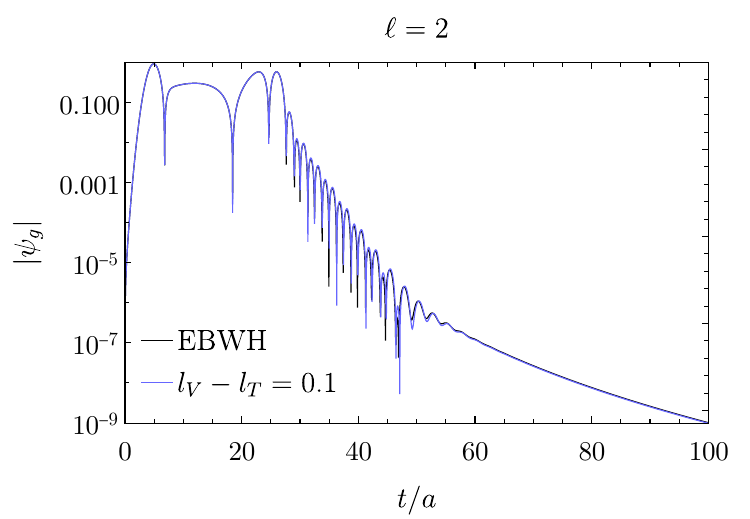}
    \caption{Temporal evolution data of scalar perturbations minimally coupled to gravity in LVEB wormholes. The black lines correspond to the perturbation of a scalar field in EBWH, that is, in the absence of LV.}
    \label{fig:time-domain_g}
\end{figure*}

In Fig.~\ref{fig:time-domain_g} we show the time-domain profiles of the scalar field perturbation minimally coupled to gravity within the LVEB wormhole for three polar indices, specifically $\ell=0,1$ and $2$, and two values of difference between LV parameters, namely $l_V-l_T=0.1$ and $-0.1$. In each panel of the figure, we compare the time-domain profile of the scalar perturbation in the LVEB wormhole with that of scalar perturbations in the EBWH (or similarly if $l_V=l_T$).

We can clearly see the three stages discussed in the previous section. At intermediate times, when the QNMs dominate, we can indeed see that the perturbations are more damped than in the EBWH for smaller values of the difference $l_V-l_T$. After the ringdown time, one appreciates the typical tail stage, with a power-law decay that appears to be independent of the LV.

We extract the QNM frequencies of the fundamental mode from the intermediate times using the Prony method and exhibit them, together with the QNM frequencies computed via DI and 6th-order WKB, in Table~\ref{tab:QNblzM_g}. As we can see, there is an excellent agreement among the three methods.

\begin{table}
\centering
\caption{Quasinormal frequencies of scalar perturbations minimally coupled to gravity in LVEB wormholes, computed via DI, 6th-order WKB and Prony methods.}
\label{tab:QNblzM_g}
\begin{tabular}{|c|c|c|c|c|}
\hline
\(\ell\) & \(l_V-l_T\) & Method & \(\text{Re}[\omega]a\) & \(-\text{Im}[\omega]a\) \\
\hline
\multirow{9}{*}{0} & \multirow{3}{*}{\(-0.1\)} & DI & $0.718331$ & $0.651175$ \\
\cline{3-5}
 & & WKB & $0.691529$  & $0.704740$ \\
\cline{3-5}
 & & Prony & $0.718678$ & $0.651793$ \\
\cline{2-5}
& \multirow{3}{*}{\(0\)} & DI & $0.681431$ & $0.617870$ \\
\cline{3-5}
 & & WKB & $0.656042$ & $0.668575$ \\
\cline{3-5}
 & & Prony & $0.681298$ & $0.617775$ \\
\cline{2-5}
& \multirow{3}{*}{\(0.1\)} & DI & $0.649596$ & $0.589163$ \\
\cline{3-5}
 & & WKB & $0.625512$ & $0.637461$ \\
\cline{3-5}
 & & Prony & $0.650444$ & $0.589188$ \\
\hline
\multirow{9}{*}{1} & \multirow{3}{*}{\(-0.1\)} & DI & $1.595446$ & $0.544144$ \\
\cline{3-5}
 & & WKB & $1.602689$  & $0.546061$ \\
\cline{3-5}
 & & Prony & $1.595667$ & $0.543993$ \\
\cline{2-5}
& \multirow{3}{*}{\(0\)} & DI & $1.572708$ & $0.529700$ \\
\cline{3-5}
 & & WKB & $1.576355$ & $0.533059$ \\
\cline{3-5}
 & & Prony & $1.572913$ & $0.529558$ \\
\cline{2-5}
& \multirow{3}{*}{\(0.1\)} & DI & $1.554710$ & $0.516504$ \\
\cline{3-5}
 & & WKB & $1.556085$ & $0.520063$ \\
\cline{3-5}
 & & Prony & $1.554904$ & $0.516367 $ \\
\hline
\multirow{9}{*}{2} & \multirow{3}{*}{\(-0.1\)} & DI & $2.564324$ & $0.519491$ \\
\cline{3-5}
 & & WKB & $2.567154$  & $0.515875$ \\
\cline{3-5}
 & & Prony & $2.564896$ & $0.519174$ \\
\cline{2-5}
& \multirow{3}{*}{\(0\)} & DI & $2.546656$ & $0.512668$ \\
\cline{3-5}
 & & WKB & $2.548319$ & $0.511121$ \\
\cline{3-5}
 & & Prony & $2.547213$ & $0.512359$ \\
\cline{2-5}
& \multirow{3}{*}{\(0.1\)} & DI & $2.533540$ & $0.504946$ \\
\cline{3-5}
 & & WKB & $2.534318$ & $0.504411$ \\
\cline{3-5}
 & & Prony & $2.534084$ & $0.504644$ \\
\hline
\end{tabular}
\end{table}

\subsubsection{Canonical scalar field coupled to the spacetime metric and to the Lorentz-violating fields}

\begin{figure*}[!htb]
    \centering
    \includegraphics[width=\columnwidth]{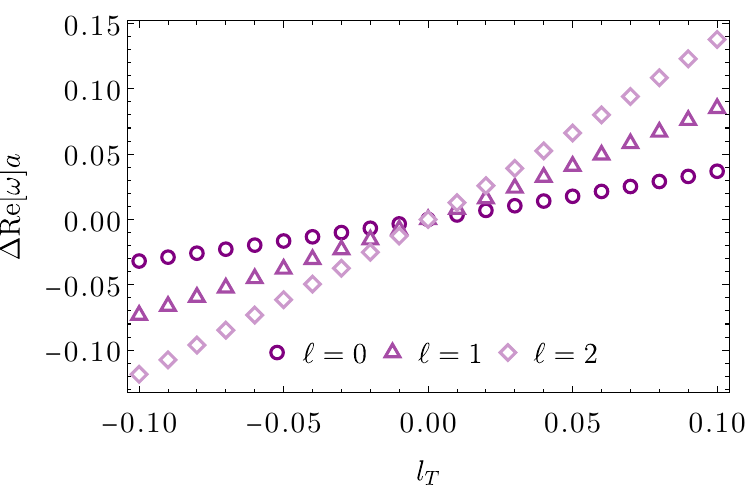} \includegraphics[width=\columnwidth]{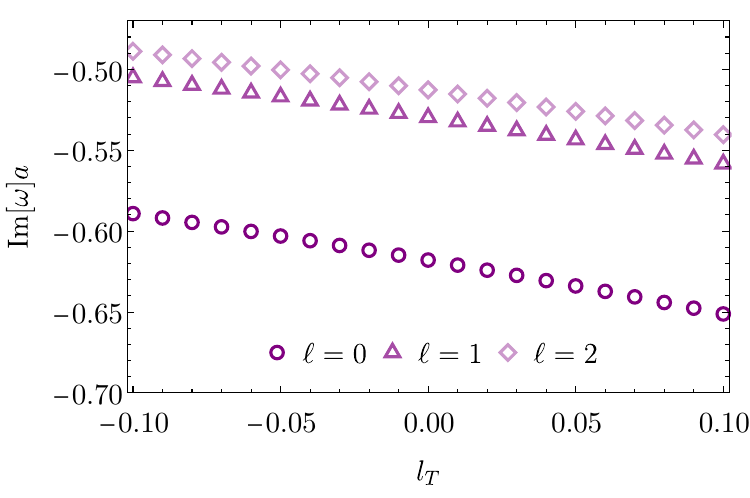}
    \caption{QNMs of scalar fields subjects to LV-matter couplings in LVEB wormholes, computed via DI. Left panel: Difference between the real part of the QNMs in LVEB wormholes and the corresponding real part in EBWHs. For $\ell=0$, $1$, and $2$, these differences are $0.68143$, $1.5727$, and $2.5466$, respectively. Right panel: Imaginary part of the QNMs.}
    \label{fig:QNM_Re_and_Im}
\end{figure*}

Now, let us consider the results for the scalar perturbation $\Psi_{LV}$ regarding that $m_V=l_V$ and $m_T=l_T$, that is, assuming that the Lorentz violation influences both the gravitational and matter sectors equally. In Fig.~\ref{fig:QNM_Re_and_Im} we show the real and the imaginary parts of the QNM frequencies, computed using DI, of scalar perturbations subjected to LV-matter couplings lying in the LVEB wormhole spacetime. In this scenario, the value of $\text{Re}[\omega]$ increases as the LV parameter $l_T$ increases. We show this behavior in the top panel of Fig.~\ref{fig:QNM_Re_and_Im} by displaying the difference $\Delta\text{Re}[\omega]=\text{Re}[\omega] -\text{Re}[\omega_0]$. Conversely, $\text{Im}[\omega]$ decreases as the LV parameter $l_T$ increases, indicating that the modes are more damped as we increase the LV parameter. We notice that, similarly to the EBWH case, the QNMs of the scalar field $\Psi_{LV}$ are always damped, thus corresponding to stable perturbations.

\begin{figure*}[!htb]
    \centering
    \includegraphics[width=\columnwidth]{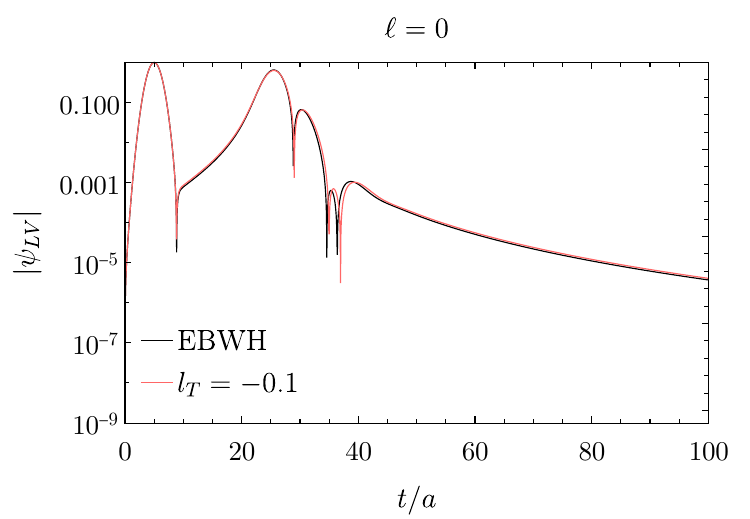}\includegraphics[width=\columnwidth]{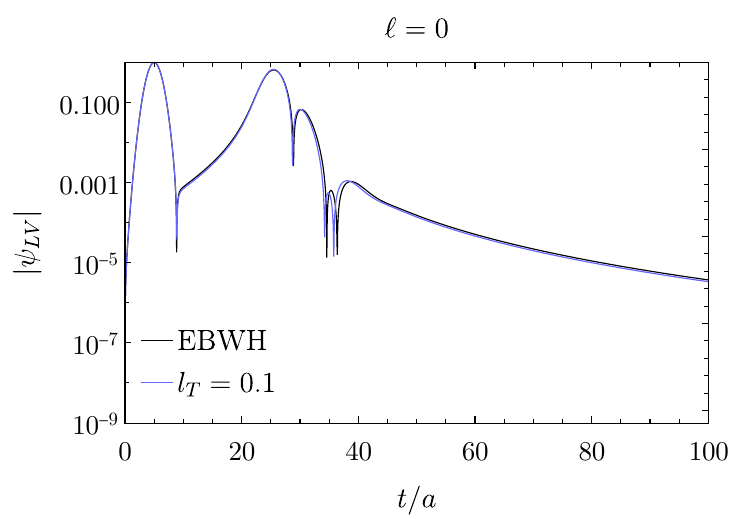}\\
    \includegraphics[width=\columnwidth]{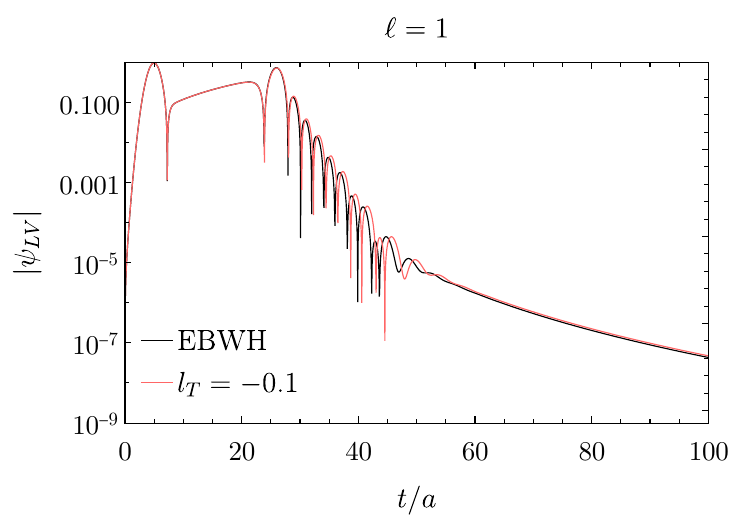}\includegraphics[width=\columnwidth]{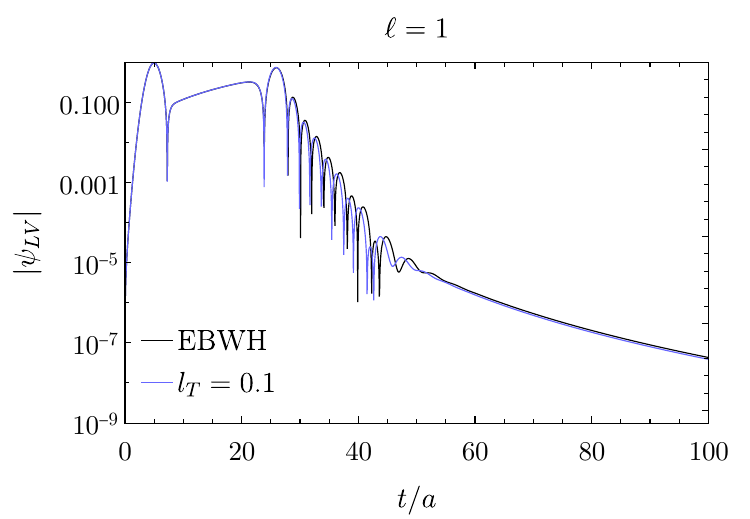}\\
    \includegraphics[width=\columnwidth]{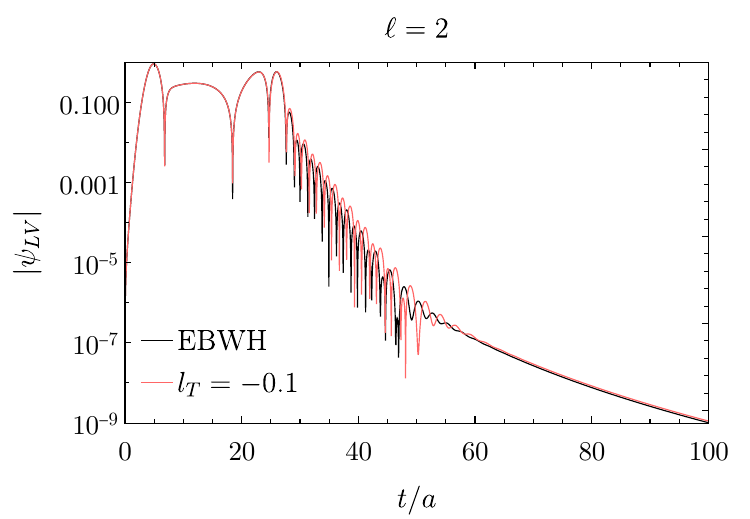}\includegraphics[width=\columnwidth]{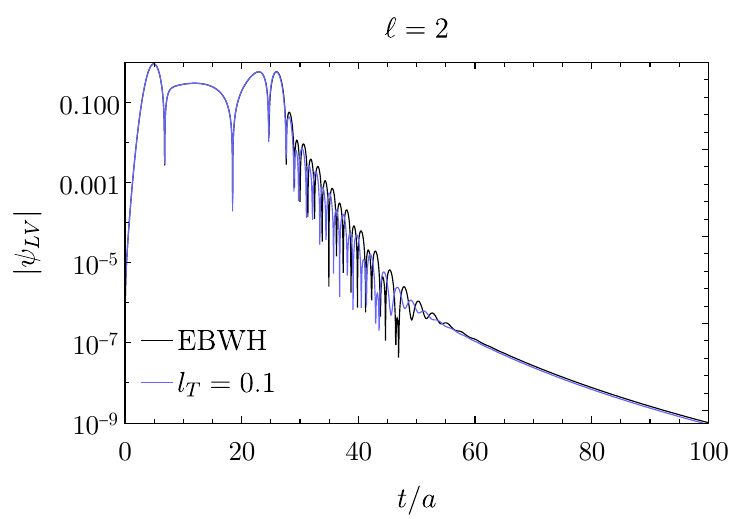}
    \caption{Temporal evolution data of scalar perturbations non-minimally coupled to Lorentz-violating fields in LVEB wormholes. The black lines correspond to the perturbation of a canonical scalar field in EBWH, that is, in the absence of LV.}
    \label{fig:time-domain}
\end{figure*}

In Fig.~\ref{fig:time-domain} we depict the time-domain profiles of the scalar field perturbation subjected to LV-matter couplings in the LVEB wormhole for three polar indices, specifically $\ell=0,1$ and $2$, and two values of LV parameter, namely $l_T=0.1$ and $-0.1$. In each panel of the figure, we compare the time-domain profile of the scalar perturbation in the LVEB wormhole with the one of scalar perturbations in EBWH. We can clearly see the three stages discussed in the aforementioned section. One observes that the perturbations are less damped for smaller LV parameter values than in the EBWH case. After the ringdown time, the power-law tail stage would appear to be independent of the LV.

\begin{table*}[!htb]
\centering
\caption{Quasinormal frequencies of scalar perturbations non-minimally coupled to Lorentz-violating fields in LVEB wormholes, computed via DI, 6th-order WKB, and Prony methods.}
\label{tab:QNblzM}
\begin{tabular}{|c|c|c|c|c|}
\hline
\(\ell\) & \(l_T\) & Method & \(\text{Re}[\omega]a\) & \(-\text{Im}[\omega]a\) \\
\hline
\multirow{9}{*}{0} & \multirow{3}{*}{\(-0.1\)} & DI & $0.649596$ & $0.589163$ \\
\cline{3-5}
 & & WKB & $0.625512$  & $0.637461$ \\
\cline{3-5}
 & & Prony & $0.650444$ & $0.589189$ \\
\cline{2-5}
& \multirow{3}{*}{\(0\)} & DI & $0.681431$ & $0.617870$ \\
\cline{3-5}
 & & WKB & $0.656042$ & $0.668575$ \\
\cline{3-5}
 & & Prony & $0.681298$ & $0.617775$ \\
\cline{2-5}
& \multirow{3}{*}{\(0.1\)} & DI & $0.718331$ & $0.651175$ \\
\cline{3-5}
 & & WKB & $0.691529$ & $0.704740$ \\
\cline{3-5}
 & & Prony & $0.718678$ & $0.651793$ \\
\hline
\multirow{9}{*}{1} & \multirow{3}{*}{\(-0.1\)} & DI & $1.499518$ & $0.505049$ \\
\cline{3-5}
 & & WKB & $1.502996$  & $0.508252$ \\
\cline{3-5}
 & & Prony & $1.499688$ & $0.504922$ \\
\cline{2-5}
& \multirow{3}{*}{\(0\)} & DI & $1.572708$ & $0.529700$ \\
\cline{3-5}
 & & WKB & $1.576355$ & $0.533059$ \\
\cline{3-5}
 & & Prony & $1.572913$ & $0.529558$ \\
\cline{2-5}
& \multirow{3}{*}{\(0.1\)} & DI & $1.657780$ & $0.558353$ \\
\cline{3-5}
 & & WKB & $1.661624$ & $0.561894$ \\
\cline{3-5}
 & & Prony & $1.658019$ & $0.558185$ \\
\hline
\multirow{9}{*}{2} & \multirow{3}{*}{\(-0.1\)} & DI & $2.428141$ & $0.488810$ \\
\cline{3-5}
 & & WKB & $2.429726$  & $0.487335$ \\
\cline{3-5}
 & & Prony & $2.428628$ & $0.488545$ \\
\cline{2-5}
& \multirow{3}{*}{\(0\)} & DI & $2.546656$ & $0.512668$ \\
\cline{3-5}
 & & WKB & $2.548319$ & $0.511121$ \\
\cline{3-5}
 & & Prony & $2.547213$ & $0.512359$ \\
\cline{2-5}
& \multirow{3}{*}{\(0.1\)} & DI & $2.684411$ & $0.540400$ \\
\cline{3-5}
 & & WKB & $2.686164$ & $0.538769$ \\
\cline{3-5}
 & & Prony & $2.685056$ & $0.540030$ \\
\hline
\end{tabular}
\end{table*}

We extract the QNM frequencies of the fundamental mode from the intermediate times using the Prony method and exhibit them, together with the QNM frequencies computed via DI and 6th-order WKB, in Table~\ref{tab:QNblzM}. As we can see, there is remarkable accord among the three methods.

\section{A note on scalar perturbations subjected to couplings to Lorentz-violating fields}\label{sec:a_note}

The intriguing fact that the scalar perturbation $\Psi_{LV}$ in LVEB wormholes may not exhibit LV effects in its dynamics when $l_T = 0$ is a direct consequence of the condition $m_V=l_V$ and $m_T=l_T$. A natural question, however, is whether this counterintuitive behavior can also occur in other scenarios with spacelike VEV configurations. We illustrate that this phenomenon similarly arises in vacuum spherically symmetric bumblebee black holes~\cite{Casana:2017jkc}, and we use this example to deduce the necessary conditions for such behavior to emerge.

First, let us consider a spherically symmetric solution of the gravity framework
\begin{align*} \label{ac2}
    S = \dfrac{1}{2\kappa}\int d^4x \sqrt{-g}&\left[R+\tilde{\xi}_2\mathfrak{B}^\mu \mathfrak{B}^\nu R_{\mu\nu} -\frac{\kappa}{2}\mathfrak{B}_{\mu\nu}\mathfrak{B}^{\mu\nu}\right. \\ &\left. -2\kappa\mathfrak{U}(\mathfrak{X})\right],\numberthis
\end{align*}
given by
\begin{equation}
    ds^2=-\left(1-\frac{2M}{r}\right)dt^2+\frac{1+l_V}{1-\dfrac{2M}{r}}dr^2+r^2d\Omega^2,
\end{equation}
with VEV configuration $\mathfrak{b}_\mu=(0,\mathfrak{b}\sqrt{1+l_V}/\sqrt{1-2M/r},0,0)$ triggered by the self-interaction term $\mathfrak{U}(\mathfrak{X})=\tfrac{1}{2}\tilde{\alpha}(\mathfrak{B}^\mu\mathfrak{B}^\nu-\mathfrak{b}^2)^2$, such that $\mathfrak{b}_\mu\mathfrak{b}^\mu=\mathfrak{b}^2$ is constant. Here, $M$ stands for the mass of the black hole and $l_V$ has the same definition of the previous sections, namely $l_V=\tilde{\xi}_2\mathfrak{b}^2$.

Now, let us additionally assume that there is a test canonical scalar field subjected to a LV-matter coupling, lying in this spherically symmetric spacetime, such that its dynamics is described by
\begin{equation}
\label{eq:scalar_h}\nabla_\mu(h^{\mu\nu}\nabla_\nu\Psi)=0,
\end{equation}
with $h^{\mu\nu}:=g^{\mu\nu}+2\tilde{\zeta} \mathfrak{b}^\mu\mathfrak{b}^\nu$, where the intensity of the LV-matter coupling is controlled by the constant $\tilde{\zeta}$. Interestingly, if one assumes that the LV-matter coupling constant is related to the LV-gravity coupling constant through $\tilde{\zeta}=\tilde{\xi}_2/2$, one obtains that $h^{\mu\nu}=g^{\mu\nu}+\tilde{\xi}_2\mathfrak{b}^\mu\mathfrak{b}^\nu$, or explicitly
\begin{widetext}
\begin{align*}
h^{\mu\nu} =&\begin{pmatrix}
-\dfrac{1}{1 - 2M/r} & 0 & 0 & 0 \\
0 & \dfrac{\left(1 - 2M/r\right)}{1+l_V} & 0 & 0 \\
0 & 0 & \dfrac{1}{r^2} & 0 \\
0 & 0 & 0 & \dfrac{1}{r^2 \sin^2\theta}
\end{pmatrix}+
\begin{pmatrix}
0 & 0 & 0 & 0 \\
0 & \dfrac{l_V\left(1 - 2M/r\right)}{1+l_V} & 0 & 0 \\
0 & 0 & 0 & 0 \\
0 & 0 & 0 & 0
\end{pmatrix}\\=&
\begin{pmatrix}
-\dfrac{1}{1 - 2M/r} & 0 & 0 & 0 \\
0 & \left(1 - 2M/r\right) & 0 & 0 \\
0 & 0 & \dfrac{1}{r^2} & 0 \\
0 & 0 & 0 & \dfrac{1}{r^2 \sin^2\theta}
\end{pmatrix}
=g_S^{\mu\nu}\numberthis
\end{align*}
\end{widetext}
where $g_S^{\mu\nu}$ are the contravariant components of the Schwarzschild metric. Therefore, under the assumption $\tilde{\zeta} =  \tilde{\xi}_2 / 2$, Eq.~\eqref{eq:scalar_h} can be written as $\nabla_\mu(g_S^{\mu\nu}\nabla_\nu\Psi)=0$. This assumption was what led, in Sec.\ref{sec:dynamics}, to the condition $m_V=l_V$. It is important to point out that the covariant derivative $\nabla$ is compatible with the spacetime metric $g_{\mu\nu}$. Still, as we shall see, it is possible to introduce a new covariant derivative, $\mathring{\nabla}$, such that it is compatible with the Schwarzschild metric, and therefore $\mathring{\nabla}_\mu\mathring{\nabla}^\mu\Psi = 0$. Consequently, the dynamics of the scalar field perturbation is absent of any trait of the LV parameter $l_V$. In order to introduce the covariant derivative compatible with the Schwarzschild metric, we recall that
\begin{equation}
    \nabla_\mu w^\nu = \mathring{\nabla}_\mu w^\nu+C^\nu{_{\mu\alpha}}w^\alpha,
\end{equation}
where $C^\nu{_{\mu\alpha}} = \Gamma^\nu{_{\mu\alpha}}-\mathring\Gamma^\nu{_{\mu\alpha}}$, with $\Gamma^\nu{_{\mu\alpha}}$ and $\mathring\Gamma^\nu{_{\mu\alpha}}$ denoting the Levi-Civita connections of the metric $g_{\mu\nu}$ and $(g_{S})_{\mu\nu}$, respectively. Thus,
\begin{align*}
\nabla_\mu(g_S^{\mu\nu}\nabla_\nu\Psi)&= \mathring\nabla_\mu(g_S^{\mu\nu}\mathring\nabla_\nu\Psi)+C^\mu{_{\mu\alpha}}g_S^{\alpha\nu}\mathring\nabla_\nu\Psi\\& =\mathring\nabla_\mu\mathring\nabla^\mu\Psi+C^\mu{_{\mu\alpha}}g_S^{\alpha\nu}\mathring\nabla_\nu\Psi,
\end{align*}
where we have used that $\nabla_\mu\Psi=\mathring\nabla_\mu\Psi$. Since $\Gamma^\mu{_{\mu\alpha}}=\partial_\alpha(\log\sqrt{-g})$ and $\mathring\Gamma^\mu{_{\mu\alpha}}=\partial_\alpha(\log\sqrt{-g_S})$, with $g_S$ being the determinant of the metric $(g_S)_{\mu\nu}$, it is straightforward to check that
\begin{equation}
\label{eq:C_vanish} C^\mu{_{\mu\alpha}}=0,
\end{equation}
and therefore the scalar field equation~\eqref{eq:scalar_h} reduces to the standard Klein-Gordon equation.

We note that the generalization of this result for scalar perturbations subjected to LV-matter couplings in general spacetimes of the system~\eqref{ac2} may not be straightforward. The spacetimes where we observed the ``masking'' effect of LV in the dynamics of (scalar) perturbations belong to a special class of solutions of the system~\eqref{ac2}, in which the bumblebee field---taken to have, at the VEV state, a constant norm and vanishing field strength---is purely radial and assumed to remain at its potential minimum. Under these particular choices of VEV states, the effect of LV appears in the spacetime metric as a factor of $1/(1+l_V)$ in the contravariant radial component \footnote{Equivalently, under a rescaling of the radial coordinate, the effect of LV appears in the spherical sector of the metric [see, for instance, Ref.~\cite{magalhaes2025wormholes}].}~\cite{lessa2025self}. Therefore, due to this simple dependence of $l_V$ on the metric and since the VEV of the bumblebee field is purely radial and $\mathfrak{b}^\mu \mathfrak{b}_\mu =\mathfrak{b}^2$, the LV-matter term contributes to attenuate (or enhance, depending on the coupling constant $\tilde{\eta}$) the factor of $l_V$ in the radial component of $h^{\mu\nu}$. Specifically, the component $h^{rr}=g^{rr}+2\tilde{\zeta} \mathfrak{b}^2 g^{rr} = g^{rr}(1+2\tilde{\zeta}\mathfrak{b}^2)$, but since, for the considered VEV configuration, $g^{rr}=A(r)/(1+l_V)$, where $A(r)$ is a function that does not depend on $l_V$, one obtains $h^{rr}=A(r)(1+2\tilde{\zeta}\mathfrak{b}^2)/(1+l_V)$. Consequently, one notices that the ``masking'' effect occurs only when $\tilde{\zeta}=\tilde{\xi}_2/2$, since this leads to $m_V\equiv 2\tilde{\zeta}\mathfrak{b}^2 =l_V$. However, we should point out that such equality of the influence of the Lorentz violation in the matter and gravitational sectors is not a universal property. Thus, in more general scenarios, the ``anisotropic'' influence of the Lorentz violation in the matter and gravitational sectors can preclude the emergence of the ``masking'' effect. Moreover, given that the condition for $h^{\mu\nu}$ to be independent of $l_V$ is necessary for the ``masking'' effect, it is expected that more general VEV configurations---for instance, those where the bumblebee field leaves its potential minimum~\cite{bailey2025bumblebee} or where its temporal and radial components vary~\cite{liu2025exact}---present technical difficulties for the emergence of the ``masking'' effect.

We remark that the generalization of the ``masking'' effect to include scenarios where an antisymmetric rank-2 tensor has a non-zero VEV, such as those considered in this paper, can be performed. By assuming that the perturbations of a (free) scalar field, say $\Psi$, is subjected to LV-matter couplings with non-vanishing VEV bumblebee, $\mathfrak{b}^\mu$, and non-vanishing VEV antisymmetric rank-2 tensor, $b^{\mu\nu}$, its dynamics is determined by
\begin{equation}
\label{eq:scalar_q_proof}\nabla_\mu(q^{\mu\nu}\nabla_\nu\Psi)=0,
\end{equation}
where $q^{\mu\nu}:=g_{LV}^{\mu\nu}+2(\tilde{\zeta}\mathfrak{b}^\mu\mathfrak{b}^\nu+\zeta b^{\mu\alpha}b^\nu{_{\alpha}})$ and $g_{LV}^{\mu\nu}$ are the contravariant components of some metric influenced by the LV induced by $\mathfrak{b}^\mu$ and $b^{\mu\nu}$. By assuming that the LV-matter couplings and the LV-gravity couplings are related by $\tilde{\zeta}=\tilde{\xi}_2/2$ and $\zeta=\xi_2/2$, one obtains $q^{\mu\nu}:=g_{LV}^{\mu\nu}+\tilde{\xi}_2 \mathfrak{b}^\mu\mathfrak{b}^\nu+\xi_2 b^{\mu\alpha}b^\nu{_{\alpha}}$. As we have seen, depending on the VEV configurations and on the structure of the spacetime metric, the auxiliary metric $q^{\mu\nu}$ can take the same form as a metric of GR, that is $q^{\mu\nu}=g_{GR}^{\mu\nu}$. Additionally, one can introduce a new covariant derivative, $\mathring\nabla$, compatible with the auxiliary metric, such that Eq.~\eqref{eq:scalar_q_proof} becomes
\begin{equation}
    \label{eq:proof_2} \mathring\nabla_\mu\mathring\nabla^\mu\Psi+C^{\mu}{_{\mu\alpha}}g_{GR}^{\alpha\nu}\mathring\nabla_\nu\Psi=0,
\end{equation}
thus the vanishing of the latter term on the left-hand side implies that the scalar perturbation propagate on the spacetime metric as it was propagating in a GR background, that is, without any trait of LV. Since $g_{GR}^{\alpha\nu}\mathring\nabla_\nu\Psi$ is, in general, non-vanishing, the condition for the scalar perturbation to satisfy the Klein-Gordon equation can be summarized as
\begin{equation}
 \label{eq:cond_KG}C^{\mu}{_{\mu\alpha}} = \partial_\alpha\left[\log\left(\dfrac{\sqrt{-g_{LV}}}{\sqrt{-q}}\right)\right]=0,
\end{equation}
where $g_{LV}$ and $q$ are the determinant of the metrics $(g_{LV})_{\mu\nu}$ and $q_{\mu\nu}$, respectively. It follows that, if $\log(\sqrt{-g_{LV}}/{\sqrt{-q}})$ is constant and $q^{\mu\nu}=g_{GR}^{\mu\nu}$, the scalar perturbations subjected to LV-matter couplings in the spacetime with LV behave identically to scalar perturbations in standard GR spacetime. Consequently, these perturbations do not experience the effects of LV.

One can check that the condition~\eqref{eq:cond_KG} is fulfilled for scalar perturbation $\Psi_{LV}$ in LVEB wormholes if $l_T=0$. In this case, under a suitable change of coordinates, namely $\rho=x/\sqrt{1+l_V}$, the contravariant components of the spacetime metric and of the auxiliary metric become, respectively,
\begin{align}
    g_{LV}^{\mu\nu}&=\text{diag}\left(-1,\frac{1}{1+l_V},\frac{1}{a^2+\rho^2},\frac{1}{(a^2+\rho^2)\sin^2\theta}\right),\\
    q^{\mu\nu}&=\text{diag}\left(-1,1,\frac{1}{a^2+\rho^2},\frac{1}{a^2+\rho^2)\sin^2\theta}\right)=g_{EB}^{\mu\nu},
\end{align}
hence $\log(\sqrt{-g_{LV}}/{\sqrt{-q}})=-\log(\sqrt{1+l_V})$ is constant, and therefore $C^{\mu}{_{\mu\alpha}}=0$. As a result, the perturbations subjected to LV-matter couplings in the LVEB wormholes with $l_T=0$ propagate as if they were lying in EBWHs.

\section{Conclusion}\label{con}
We have obtained traversable wormhole solutions supported by phantom scalar fields within a Lorentz-violating gravity framework by incorporating non-trivial couplings in the matter sector. Specifically, we consider a unified framework in which both the bumblebee, a vector field, and the Kalb-Ramond, a rank-2 tensor field, are coupled to the Ricci tensor and their non-zero VEVs directly influence the dynamics of the scalar fields through non-minimal couplings in the matter sector. The solutions presented here fill a gap in the search for self-consistent non-vacuum solutions in Lorentz-violating gravity models with non-minimal couplings to the Ricci tensor. Surprisingly, the wormhole solutions obtained in this work exhibit a structure remarkably similar to the LVEB wormholes previously found in the literature, which were derived considering non-minimal couplings to the Riemann tensor~\cite{magalhaes2025wormholes}. However, the models presented here are generated by scalar fields without a self-interaction potential.

Remarkably, the effective contributions of LV from the non-zero VEV of the bumblebee field and the non-zero VEV of the antisymmetric rank-2 tensor are additive, despite originating from fields with different natures ($\mathfrak{b}^\mu$ has a spacelike norm whereas $b^{\mu\nu}$ has a timelike norm). Notably, when the LV parameters satisfy a specific configuration, namely $l_V = l_T$, the line element of the LVEB wormhole (see Eq.~\eqref{eq:ellis_wormholemixed}) simplifies to that of the EB wormhole. Consequently, experiments that focus solely on spacetime curvature effects may fail to detect LV signatures. Further investigation into these gravity frameworks, particularly with two tensor fields that spontaneously break Lorentz symmetry, is essential to determine whether more astrophysically relevant solutions, such as black holes, can accommodate distinct combinations of LV parameters.

A promising avenue for detecting imprints of LV lies in analyzing the perturbations within these spacetimes. Since the Lorentz-violating fields can, in principle, couple to independent canonical scalar fields, the investigation of how the dynamics of such scalar fields are influenced by LV effects and how their dynamics can be different from the dynamics of scalar fields minimally coupled to the spacetime metric is well-motivated. The role of LV parameters in these scenarios can differ notably, so we expect distinct QNM spectra. We therefore computed these spectra using three distinct methods: the DI method, the 6th-order WKB approximation, and the Prony method, finding excellent agreement among the results.

Concerning scalar fields minimally coupled to the spacetime metric, our analysis for the LVEB wormholes reveals that the real part of the QNM frequencies decreases as the difference $l_V-l_T$ increases, whilst the imaginary part increases as the difference increases. Therefore, the numerical results expose that the scalar field perturbations are more damped at smaller values of $l_V-l_T$. On the other hand, the analysis of scalar fields subject to LV-matter couplings is subtler. We have focused on the case where the influence of the Lorentz violation is the same in both matter and gravitational sectors, which is achieved by the conditions $m_V=l_V$ and $m_T=l_T$. Scalar fields coupled to a stationary spacelike background vector (bumblebee) field propagate in a spacetime where the same vector spontaneously triggers LV as they would in a GR background. We illustrate this intriguing behavior for perturbations in the wormhole spacetimes with $l_T=0$. Correspondingly, the QNM spectra of these scalar perturbations match those of the EB wormhole. Similarly, this also happens in the static, spherically symmetric vacuum black holes found in Einstein-bumblebee gravity~\cite{Casana:2017jkc}. Conversely, when $l_T\neq 0$, the scalar perturbations are influenced by the Lorentz violation when coupled to Lorentz-violating fields. In these scenarios, the non-vanishing VEV of the antisymmetric rank-2 tensor alters the QNM spectra compared to the EB wormhole. Our analysis reveals that the real part of the QNM frequencies increases with $l_T$, while the imaginary part decreases (becoming more negative) as $l_T$ increases. Consequently, the scalar field perturbations are more damped at higher values of $l_T$.

Our results demonstrate that the inclusion of couplings between matter fields and tensor fields that spontaneously trigger Lorentz violation can introduce new gravitational physics. Such new effects may manifest either through the emergence of UCOs or through modifications to the dynamical equations governing perturbations. The latter is particularly intriguing, as the anisotropies induced by LV may affect the radial stability of compact objects supported by matter distributions coupled to Lorentz-violating tensor fields. A detailed analysis of the radial stability of LVEB wormholes is currently underway. We also note that an important direction for future work is to extend the analysis of perturbations subjected to LV-matter couplings to cases where $m_V \neq l_V$ and $m_T \neq l_T$, as such ``anisotropies'' between matter and gravitational sectors may reveal novel phenomenological signatures and further constrain the viability of Lorentz-violating scenarios.

\appendix
\section{Methods to compute QNMs}\label{app:methods}

In order to determine the QNMs associated with scalar perturbations, several methods are available in the literature, either in the time or frequency domains. Here we present the methods we use in our analysis.

\subsection{Time domain integration and Prony method}

Without assuming the temporal dependence $\exp(-i\omega t)$ in the decomposition of the scalar perturbations~\eqref{eq:scalarfield_decomp1} and \eqref{eq:scalarfield_decomp2}, the equation for the perturbation becomes
\begin{align}
	\label{eq:phi_t}
	\left(\dfrac{\partial^2}{\partial t^2}-\dfrac{\partial ^2}{\partial x^2}+V_{\ell}\right)\psi(t,x)=0,
\end{align}
where $V_{\ell}$ for $\psi_g$ and $\psi_{LV}$ is given by
\begin{align*}
	\label{eq:Vphi} V_{\ell}^g &= \frac{(1+l_V-l_T)(1+\ell(\ell+1)(1+l_V-l_T))a^2}{\left(a^2(1+l_V-l_T)+x^2\right)^2}\\&+\frac{\ell(\ell+1)x^2}{\left(a^2(1+l_V-l_T)+x^2\right)^2},\numberthis \\
    V_{\ell}^{LV} &= \frac{-a^2 (l_T-1) \left(\ell^2+\ell+1\right)+\ell (\ell+1) x^2}{\left(x^2-a^2 (l_T -1)\right)^2},\numberthis
\end{align*}
respectively. Equation~\eqref{eq:phi_t} can be integrated following the Gundlach-Price-Pullin discretization scheme~\cite{gundlach1994late}, which involves introducing light-cone coordinates, specifically $v\equiv t+x$ and $u\equiv t-x$. Thus, the wave equation can be expressed as
\begin{equation}\label{eq:uvdiffeq}
	\left(4\frac{\partial^2}{\partial u\partial v}+V_\ell\right)\psi=0.
\end{equation}
The numerical integration is then performed on a null grid, leading to the following expression for the discretized evolution of the scalar field 
\begin{equation}\label{eq:discretizedPhi}
	\psi_N=\psi_E+\psi_W-\psi_S-\frac{h^2}{8} V_\ell(S)(\psi_W+\psi_E) + O(h^4),
\end{equation}
where $h$ is the stepsize between two neighboring grid points, and subscripts indicate the point in the grid where the function  $\psi$ is evaluated, explicitly, in $S=(u,v)$, $W=(u+h,v)$, $E=(u,v+h)$ and $N=(u+h,v+h)$. As initial conditions for the scalar perturbation, we use a Gaussian distribution on the $u=0$ surface, together with a constant profile on the $v=0$ surface, i.e.,
\begin{equation}
	\psi(0,v)=\mathcal{I}e^{-(v-v_c)^2/2\sigma^2},
\end{equation}
with height $\mathcal{I}=1$, width $\sigma^2=1$, and centered at $v_c=5$. The temporal evolution data can be divided into three distinct stages as follows: (i) an initial prompt response, heavily influenced by the selected initial conditions of the field; (ii) an exponential decay during intermediate times, governed by the QNMs; and (iii) a power-law decline at late times, resulting from the backscattering of the field in the tail of the potential~\cite{krivanDynamicsScalarFields1996}.

With the aim to extracting the QNM frequencies from the time-profile data, we use the Prony method~\cite{berti2007mining}. This method fits data using a linear combination of exponentials. Thus, the temporal evolution data can be expanded as
\begin{equation}
    \psi(t)=\sum_{j=0}^p C_j \exp(-i\omega_j t),
\end{equation}
where $C_j$ are complex coefficients, from which one solely considers the damped oscillation part of the signal, say from $t_0$ to $t=t_0+2p\Delta t$, where $\Delta t$ is the time interval of each point, and $2p\in\mathbb{Z}$ is the number of sample signals. Hence, the temporal evolution data can be written as
\begin{equation}
    \label{eq:sample}\psi_n = \sum_{j=0}^p D_j z_j^n,
\end{equation}
where $\psi_n=\psi(t_0+nh)$, $D_j=\exp(-i\omega t_0)C_j$, and $z_j=\exp(-i\omega_j\Delta t)$. The core of the Prony method is to recognize that the above equation is the solution to some homogeneous linear constant-coefficient difference equation~\cite{berti2007mining}. Thus, it is convenient to introduce the polynomial
\begin{equation}
   \label{eq:poly} P(z) = \prod_{k=1}^p(z-z_k)=\sum_{j=0}^p\alpha_i z^{p-j},
\end{equation}
which, by construction, has the $z_k$ as its roots. Since $\forall \, j, \,1\leq j\leq p$, $P(z_j)=0$, hence it follows that
\begin{equation}
    \sum_{i=0}^p\alpha_i\psi_{j-i}=\sum_{i=0}^p\alpha_i\sum_{k=1}^p D_kz_k^{j-i}=\sum_{k=1}^pD_kz_k^{j-p}(z_k)=0.
\end{equation}
By taking $\alpha_0=1$, the above equation becomes
\begin{equation}
    \sum_{k=1}^p\alpha_k\psi_{j-k}=-\psi_j.
\end{equation}
Taking $j$ from $p+1$ to $2p$ yields $p$ equations, and so we can determine the $\alpha_k$ coefficients. Once these coefficients are achieved, we can obtain the roots $z_k$ by substituting $\alpha_k$ into Eq.~\eqref{eq:poly}. The quasinormal frequencies are therefore obtained by the relation $\omega_k=\frac{i}{\Delta t}\log(z_k)$, while the amplitudes $D_j$ are attained by inverting Eq.~\eqref{eq:sample}.

\subsection{Direct integration}
In order to determine the QNM frequencies, one can solve the radial equations for the scalar perturbations directly---namely, 	 Eqs.~\eqref{eq:caseIII_phatom_pert} and~\eqref{eq:caseIII_phatom_pert_2}, for $\psi_g$ and $\psi_{LV}$, respectively. The DI is based on the shooting method and numerical root finding to obtain frequencies in the complex domain~\cite{konoplya2011quasinormal}. We divide the space at some value $-\infty<x_m<\infty$, then we integrate the corresponding radial equation from plus (minus) infinity backward (inward) to $x_m$, subjected to the corresponding boundary conditions, obtaining two solutions, respectively, $\psi_{\pm}$. The condition for the quasinormal modes is the vanishing of the Wronskian of the two solutions in the matching point, namely
\begin{equation}
    (\psi_+\psi'_- - \psi_-\psi'_+)\vert_{x_m}=0.
\end{equation}
The quasinormal frequencies are the roots of the above condition, where one can use standard root-finding algorithm, such as Newton's method.

To enhance the convergence, we write the solution of the radial equation at both infinities as a power series, respectively
\begin{align}
    \label{eq:u_series}\psi(x\to\pm\infty) = \exp(\pm i\omega x)\sum_{i=0}^N\frac{h^{\pm}_i}{x^i},
\end{align}
where $h_i^{\pm}$ are the expansion coefficients and $N$ is the order of the expansion. The coefficients are determined by plugging Eq.~\eqref{eq:u_series} in the differential equation for the radial part of the corresponding scalar field. We use Eq.~\eqref{eq:u_series} and its first-order derivative as boundary conditions to integrate Eqs.~\eqref{eq:caseIII_phatom_pert} and~\eqref{eq:caseIII_phatom_pert_2}, for $\psi_g$ and $\psi_{LV}$, respectively. In our numerical calculations, without loss of generality, we have set $h_0^\pm=1$ and $N=20$.

\subsection{WKB}

Schr\"odinger-like equations, such as Eq.~\eqref{eq:schrodinger-like}, are suitably solved using the Wentzel-Kramers-Brillouin (WKB) method if the effective potential $U_{\omega \ell}$ has a single peak, located at $x=x_0$, and approaches constant (negative) values as $x\to\pm\infty$~\cite{konoplyaHigherOrderWKB2019}, which is indeed the case of the potential of the scalar perturbations studied here. The WKB method is based on the matching of the solutions of Eq.~\eqref{eq:schrodinger-like} as $x\to\pm\infty$, with the Taylor expansion around the peak of the effective potential at the zeroes of $U_{\omega\ell}$ (turning points). From this approach, one achieves the WKB rule
\begin{equation}
    \frac{iU_{\omega\ell}(x_0)}{\sqrt{2U_{\omega\ell}^{\prime\prime}(x_0)}}-\sum_{j=2}^N \Lambda_j=n+\frac{1}{2},
\end{equation}
where $n$ labels the overtones, with $n=0$ being the fundamental mode, and the $\Lambda_j$ are corrections terms that depends on higher-order derivatives of $U_{\omega\ell}$ evaluated at $x_0$. The first-order correction was derived in Ref.~\cite{schutzBlackHoleNormal1985}, while higher-order corrections were progressively developed: up to third order in Ref.~\cite{iyerBlackholeNormalModes1987}, up to sixth order in Ref.~\cite{konoplyaQuasinormalBehaviorDimensional2003}, and up to thirteenth order in Ref.~\cite{matyjasekQuasinormalModesBlack2017}. In this work, to validate our numerical results, we use a 6th-order WKB approximation and find excellent agreement with the previous two methods.

\begin{acknowledgments}
The authors would like to acknowledge Fundação de Amparo à Pesquisa e ao Desenvolvimento Científico e Tecnológico do Maranhão (FAPEMA),  Conselho Nacional de Desenvolvimento Cient\'ifico e Tecnol\'ogico (CNPq), Coordena\c{c}\~ao de Aperfei\c{c}oamento de Pessoal de N\'ivel Superior (CAPES) -- Finance Code 001, from Brazil, for partial financial support. R.B.M. is supported by CNPq/PDJ 151250/2024-3.  L.A.L is supported by FAPEMA BPD- 08975/24. R. C. acknowledges the support from the grants CNPq/312155/2023-9 and FAPEMA/UNIVERSAL-00812/19.
\end{acknowledgments}

\bibliography{refs.bib}
\bibliographystyle{report}
\end{document}